\documentclass[acmsmall,screen,nonacm]{acmart}
\AtBeginDocument{%
  }

\usepackage{fontspec}
\usepackage{graphicx}
\usepackage{etoolbox}
\makeatletter
\patchcmd{\NAT@test}{\else \NAT@nm}{\else \NAT@hyper@{\NAT@nm}}{}{}
\makeatother

\usepackage{listings}
\usepackage{booktabs}
\usepackage{paralist}

\usepackage{hyperref}
\hypersetup{
  hidelinks,
  colorlinks=true,
  allcolors=black,
  pdfstartview=Fit,
  breaklinks=true
}

\usepackage[all]{hypcap}
\usepackage[caption=false,font=footnotesize]{subfig}
\usepackage{xspace}
\usepackage{subcaption}
\usepackage{collcell}
\usepackage{mathtools}
\usepackage{mdframed}
\usepackage{amsmath}
\usepackage{varioref}
\usepackage{scalerel}
\usepackage{array,varwidth}
\usepackage{makecell}
\usepackage{diagbox}
\usepackage{float}
\usepackage{xfrac}
\usepackage{enumitem}
\usepackage{nicematrix}
\usepackage{cleveref}
\usepackage{stmaryrd}
\usepackage[export]{adjustbox}
\usepackage{tikz}
\usepackage{dsfont}
\usepackage{bbold}
\usepackage{fbox}
\usepackage{bcprules}
\usepackage[dvipsnames]{xcolor}
\usepackage{wrapfig}
\usepackage{pifont}

\usepackage{color}

\definecolor{light-gray}{gray}{0.9}
\definecolor{dark-gray}{gray}{0.4}
\definecolor{ckeyword}{HTML}{000000}
\definecolor{ccomment}{HTML}{636363}
\definecolor{cstring}{HTML}{2A0099}

\definecolor{myyellow}{HTML}{DFE674}
\definecolor{myblue}{HTML}{F49EA0}

\colorlet{shadecolor}{light-gray!40}

\lstdefinelanguage{Scala}%
{morekeywords={
  abstract, sealed, lazy,
  case,catch,char,class,%
  def,do,else,extends,final,finally,for,%
  if,import,implicit,%
  match,module,%
  new,null,undefined,%
  array,
  override,%
  package,private,protected,public,%
  for,public,return,super,%
  this,throw,trait,try,type,%
  val,var,%
  with,while,%
  object,
  let,skip,assert,then,fst,snd,idx,sum,prod,exists,forall,%
  yield,%
  Expr,Int,Unit,Boolean,Ref,Code
  },%
  sensitive,%
  moredelim=*[il][\bfseries]{\#\#\ },
  morecomment=[l]//,%
  morecomment=[s]{/*}{*/},%
  morestring=[b]",%
  showstringspaces=false%
}[keywords,comments,strings]%

\lstdefinelanguage{OCaml}%
{morekeywords={
  abstract, sealed, lazy,
  case,catch,char,class,%
  def,do,else,extends,final,finally,for,%
  if,import,implicit,%
  match,module,%
  new,null,undefined,%
  array,
  override,%
  package,private,protected,public,%
  for,public,return,super,%
  this,throw,trait,try,type,%
  val,var,%
  with,while,%
  object,
  let,skip,assert,then,fst,snd,idx,sum,prod,exists,forall,%
  yield,%
  Expr,Int,Unit,Boolean,Ref,Code,
  ref, run, lift, fun, in, rec, code, int, genlet
  },%
  sensitive,%
  moredelim=*[il][\bfseries]{\#\#\ },
  morecomment=[s]{(*}{*)},%
  escapeinside={(*@}{@*)},
  morestring=[b]",%
  showstringspaces=false%
}[keywords,comments,strings]%

\newcommand{\commentstyle}[1]{\color{ccomment}\itshape{#1}}
\newcommand{\keywordstyle}[1]{\color{ckeyword}\bfseries{#1}}
\newcommand{\stringstyle}[1]{\color{cstring}\text{#1}}

\definecolor{listingbg}{RGB}{240, 240, 240}

\lstnewenvironment{listing}{\lstset{language=Caml}}{}
\newcommand{\code}[1]{\lstinline[language=Caml,columns=fixed,basicstyle=\ttfamily]|#1|}

\newcommand{\silent}[1]{}

\newcommand{\Keywd}[1]{\ensuremath{\mathsf{#1}}}

\newcommand{\eg}{\emph{e.g.}\xspace}
\newcommand{\ie}{\emph{i.e.}\xspace}

\newlength{\trulemargin}
\newlength{\trulewidth}
\newlength{\srulewidth}
\newenvironment{trules}{$\vspace{0.5em}\ba{p{\trulemargin}@{~}p{\trulewidth}@{~}p{\trulemargin}}}{\ea$}
\newenvironment{srules}{$\vspace{0.5em}\ba{p{\trulemargin}@{~}p{\srulewidth}}}{\ea$}

\newcommand{\ba}{\begin{array}}
\newcommand{\ea}{\end{array}}

\newcommand{\ei}{\end{array}}
\newcommand{\bcases}{\left\{\begin{array}{ll}}
\newcommand{\ecases}{\end{array}\right.}

\newcommand{\judgement}[2]{{\textsf{\textbf{#1}}} \hfill #2}

\newcommand{\BOX}[1]{\fbox{$\strut #1$}}

\newcommand{\WF}{\ensuremath{\mathsf{WF}}}

\newcommand{\lang}{\ensuremath{\lambda_{\mid\mathsf{2}|}}\xspace}
\newcommand{\langref}{\ensuremath{\lambda^\mathsf{ref}_{|\mathsf{2}|}}\xspace}

\newcommand{\langsnd}{\ensuremath{\lambda^{\shortdownarrow}_{|\mathsf{2}|}}\xspace}
\newcommand{\langsndref}{\ensuremath{\lambda^{\mathsf{ref}\shortdownarrow}_{|\mathsf{2}|}}\xspace}

\newcommand{\tmrun}[1]{\ensuremath{\mathsf{run}~#1}}
\newcommand{\tmlet}[3]{\ensuremath{\mathsf{let}~#1=#2~\mathsf{in}~#3}}
\newcommand{\tmletc}[3]{\ensuremath{\mathsf{let_c}~#1=#2~\mathsf{in}~#3}}
\newcommand{\tmlift}[1]{\ensuremath{\mathsf{lift}~#1}}
\newcommand{\tmlamc}[2]{\ensuremath{\lambda_{\mathsf{c}}#1.#2}}
\newcommand{\tmcode}[1]{\ensuremath{\mathsf{code}~#1}}
\newcommand{\tmreflect}[1]{\ensuremath{\mathsf{reflect}~#1}}
\newcommand{\tmappSt}[2]{\ensuremath{\mathsf{app}^\mathbb{1} ~ {#1} ~ {#2}}}
\newcommand{\tmappDy}[2]{\ensuremath{\mathsf{app}^\mathbb{2} ~ {#1} ~ {#2}}}
\newcommand{\tmfixSt}[1]{\ensuremath{\mathsf{fix}^\mathbb{1} ~ {#1}}}
\newcommand{\tmfixDy}[1]{\ensuremath{\mathsf{fix}^\mathbb{2} ~ {#1}}}
\newcommand{\tmifzSt}[3]{\ensuremath{\mathsf{ifz}^\mathbb{1} ~ {#1} ~ {#2} ~ {#3}}}
\newcommand{\tmifzDy}[3]{\ensuremath{\mathsf{ifz}^\mathbb{2} ~ {#1} ~ {#2} ~ {#3}}}
\newcommand{\tmallocSt}[1]{\ensuremath{\mathsf{alloc}^\mathbb{1} ~ {#1}}}
\newcommand{\tmallocDy}[1]{\ensuremath{\mathsf{alloc}^\mathbb{2} ~ {#1}}}
\newcommand{\tmgetSt}[1]{\ensuremath{\mathsf{get}^\mathbb{1} ~ {#1}}}
\newcommand{\tmgetDy}[1]{\ensuremath{\mathsf{get}^\mathbb{2} ~ {#1}}}
\newcommand{\tmputSt}[2]{\ensuremath{\mathsf{put}^\mathbb{1} ~ {#1} ~ {#2}}}
\newcommand{\tmputDy}[2]{\ensuremath{\mathsf{put}^\mathbb{2} ~ {#1} ~ {#2}}}

\newcommand{\tyrep}[1]{\ensuremath{\mathsf{rep}~#1}}
\newcommand{\tyref}[1]{\ensuremath{\mathsf{ref}~#1}}
\newcommand{\tyfrag}[1]{\ensuremath{\mathsf{frag}~#1}}
\newcommand{\vgap}{\vspace{5pt}}

\newcommand{\headto}{\ensuremath{\leadsto}}

\newcommand{\steprule}[1]{{\color{dark-gray}\ensuremath{\textsc{(#1)}}}}

\makeatletter
\newcommand\dotover{\leavevmode\cleaders\hb@xt@ .22em{\hss $\cdot$\hss}\hfill\kern\z@}
\newcommand{\dotfrac}[2]{
\ooalign{$\genfrac{}{}{0pt}{0}{#1}{#2}$\cr\dotover\cr}
}
\makeatother

\begin{document}

\title[When Do Staging Annotations Preserve Semantics? (Extended Version)]{When Do Staging Annotations Preserve Semantics? Mechanizing Typed Semantics-Preserving Multi-stage Programming with Let-Insertion (Extended Version)}

\author{Jun Tan}
\orcid{0009-0001-8915-3753}
\affiliation{%
  \institution{Independent}
  \city{Shanghai}
  \country{China}
}
\email{tjapzjj@gmail.com}

\author{Guannan Wei}
\correspondingauthor
\orcid{0000-0002-3150-2033}
\affiliation{%
  \institution{Tufts University}
  \city{Medford}
  \country{USA}
}
\email{guannan.wei@tufts.edu}

\renewcommand{\shortauthors}{Tan and Wei}

\begin{abstract}
  Multi-stage programming with quotations has long provided a powerful way
  to generate and manipulate code.
  By treating code as data, programmers can write multi-stage programs in which
  earlier stages produce specialized code from inputs available at
  generation time.
  Modern typed multi-stage languages (\eg, MetaML, MetaOCaml, Template Haskell,
  and Scala 3) adopt quotation/splicing constructs while enforcing the
  well-typedness of generated code.
  However, manipulating code fragments syntactically can subtly change evaluation
  order, leading to semantic discrepancies between a staged program and its
  unstaged counterpart, which is intended to serve as a reference
  implementation in many cases.
  The inconsistency complicates reasoning about correctness, and prevents staged
  code from being a drop-in replacement for its unstaged counterpart.

  In this paper, we study the design of multi-stage languages with semantics
  preservation guarantees.
  We develop two statically typed two-stage calculi, \lang and \langref, the
  latter supporting mutable references in the second stage.
  Their dynamic semantics model automatic let-insertion, tracked as a control
  effect in a lightweight type-and-effect system, enabling type-safe and
  semantics-preserving manipulation of effectful code fragments.
  We develop binary logical relations to prove strong semantics-preservation theorems:
  if a well-typed two-stage
  program $t_1$ evaluates to a value $\tmcode{t_2}$, then $t_2$ is contextually
  equivalent to the stage-erasure of $t_1$.
  Our calculi and their mechanized metatheory provide a simple and definitive
  answer to the question posed by
  \citet{DBLP:conf/esop/InoueT12,DBLP:journals/jfp/InoueT16} of when staging
  annotations preserve semantics, and lay a foundation for future work on
  semantics-preserving multi-stage programming.
\end{abstract}

\keywords{staging, let-insertion, partial evaluation, type systems, mechanization}

\maketitle

\section{Introduction}

Multi-stage programming (MSP) with quoting and splicing has long provided a
powerful means for code generation and manipulation, from the early days of
Lisp \cite{DBLP:conf/pepm/Bawden99} to MetaML
\cite{DBLP:conf/pepm/TahaS97, taha1999multistage}
and its descendants. By constructing the meta-program
with quotations and splices, MSP allows programs to execute in
multiple stages, where earlier stages generate code specialized to the data
available at those stages.
Staging annotations provide a natural and programmer-controlled
form of partial evaluation \cite{DBLP:books/daglib/0072559},
as well as an effective mechanism for domain-specific program optimization.

With the syntactic form of annotations,
MetaML \cite{DBLP:conf/pepm/TahaS97, taha1999multistage} further ensures hygiene and
type safety of generated code by type systems, giving rise to what is now
generally referred to as multi-stage programming.
Since the introduction of MetaML, typed MSP with syntactic annotations has grown into
a large family of research calculi (\eg, \cite{DBLP:conf/gpce/YaguchiK25,
DBLP:journals/pacmpl/XieWNY23, 10.1145/1809028.1806642}) and has been adopted in various industrial languages,
including
MetaOCaml \cite{DBLP:journals/ftpl/Kiselyov18, DBLP:conf/flops/Kiselyov14, DBLP:journals/scp/Kiselyov26},
F\# \cite{DBLP:journals/pacmpl/Syme20},
(Typed) Template Haskell \cite{DBLP:journals/pacmpl/XiePLWYW22, DBLP:journals/sigplan/SheardJ02},
and more recently Scala 3 \cite{DBLP:conf/gpce/StuckiBO18, DBLP:conf/gpce/StuckiBO21}.

\subsubsection*{\textbf{Semantics-Breaking Staging Annotations}}
Despite the expressive power and popularity of MetaML-style syntactic MSP, multi-stage programs
are notoriously difficult to reason about \cite{DBLP:journals/jfp/InoueT16,
DBLP:conf/esop/InoueT12}.
Unlike the name ``annotations'' might suggest, annotations are not just
syntactic markers that can be freely added or removed without changing the
semantics of the program.
They can rearrange the
evaluation order within the same stage, which can lead to inconsistent behaviors between
the staged program and its original unstaged counterpart (obtained by erasing
staging annotations) in the presence of computational effects, such as
divergence and mutable states.

We use the syntax from MetaOCaml \cite{DBLP:journals/scp/Kiselyov26} to
illustrate a few examples.  Brackets \code{.<}$e$\code{>.} delay the
evaluation of the enclosed expression $e$ as a code value, and an
escape \code{.~}$e$ splices the value of $e$ into a surrounding code fragment.
The type $\tau$~\code{code} denotes code fragments that, when run, produces a value of type $\tau$.
Now let us consider the following \emph{first-stage} function \code{k} (adapted from the SKI
combinators), which takes two \emph{second-stage} integers and returns the first one:
\begin{lstlisting}[language=OCaml]
  let k (x : int code) (y : int code) : int code = x in
  let rec loop () = loop () in
  k .<42>. .<loop ()>.
\end{lstlisting}
Here, when applying \code{k}, the first argument is a code fragment of integer
\code{.<42>.}, and the second is a code fragment of a non-terminating computation
\code{.<loop ()>.}.
The first-stage evaluation terminates and produces a \emph{code} \code{.<42>.}, which can be
further evaluated to \emph{integer} 42 when run.
Since \code{k} simply returns the first argument, quoting not only delays the
evaluation of \code{loop ()}, but completely prevents it from being evaluated
\emph{at all}.

However, if we compare the staged program with its unstaged counterpart, we can
see that they behave differently.
The unstaged version obtained by erasing staging annotations (brackets, escapes,
and $\tau$~\code{code} types) is the following, which diverges under call-by-value (CBV):
\begin{lstlisting}[language=OCaml]
  let k (x : int) (y : int) : int = x in
  let rec loop () = loop () in
  k 42 (loop ())
\end{lstlisting}
Moreover, splicing code fragments at will can lead to code duplication, and with
it, duplicated effects.
Consider the following \code{twice} function that splices the argument
\code{x} twice into the body, effectively adding the argument to itself:
\begin{lstlisting}[language=OCaml]
  (* erased: let twice x = x + x *)
  let twice (x : int code) : int code = .<.~x + .~x>. in ...
\end{lstlisting}
Now invoking \code{twice} with a code fragment inducing the second-stage stdout effect:
\begin{lstlisting}[language=OCaml]
  twice .<(print_string "Hello"; 42)>.
  (* generates code by copying the argument twice: *)
  .<(print_string "Hello"; 42) + (print_string "Hello"; 42)>.
\end{lstlisting}
The staged program generates code that prints twice.
In contrast, the unstaged (or annotation-erased) program evaluates the argument
only once \emph{before} the function call and thus prints \code{"Hello"} only once,
consistent with standard CBV evaluation.

\subsubsection*{\textbf{Why it Matters?}}
These examples show that, because syntactic staging annotations
permit evaluation to be rearranged, they can arbitrarily discard, reorder, or
duplicate effects.
Although such a flexibility can be useful for program manipulation in some cases,
we lose the ability to relate the staged program with the unstaged program,
thus weakening semantic predictability and making staged
programs much harder to reason about.

This limitation particularly hinders performance-oriented uses of staging,
such as Futamura projections \cite{futamura1971partial} and domain-specific
staging \cite{DBLP:conf/sigmod/TahboubER18, DBLP:conf/nips/WangDWER18,
DBLP:journals/pacmpl/WeiCR19, DBLP:conf/icse/WeiJGDTBR23,
DBLP:journals/pacmpl/BallantyneSHBA25},
whose \emph{goal is precisely to produce faster program without changing the
observable behavior of the original unstaged program}.
If staging annotations do not preserve semantics, the unstaged program cannot
serve as a reference specification from which one derives a faster, equivalent
version by merely inserting staging annotations.  Therefore, the generated
program cannot be used as drop-in replacement for the original.

The key property connecting a staged program and its unstaged
counterpart is known as \emph{erasure soundness} \cite{DBLP:journals/jfp/InoueT16,
DBLP:conf/esop/InoueT12, yang2000reasoning} or semantics preservation.
Although it was recognized as an important property
\cite{DBLP:journals/jfp/InoueT16, DBLP:conf/esop/InoueT12},
MetaML-style CBV staging unfortunately does not enjoy this property in general.
In fact, staging in CBN calculi is better behaved, since the semantics of
quotation remains aligned with the underlying CBN semantics
\cite{DBLP:conf/esop/InoueT12, DBLP:journals/jfp/InoueT16, taha1999multistage}.
However, adding staging annotations to a CBV language breaks the host language's
equational theory \cite{DBLP:conf/esop/InoueT12, DBLP:journals/jfp/InoueT16}.
As a result, the standard reasoning principles for CBV programs can no longer be
assumed to apply to staged programs.

Semantic preservation is also the cornerstone of automated partial
evaluation~\cite{DBLP:books/daglib/0072559}. As its successor, MSP gives
programmers explicit control over binding-times but abandons guarantees that
the generated code preserves the original program's semantics.

\subsubsection*{\textbf{``When Do We Know That Erasing Annotations Preserves Semantics?''}}
Nevertheless, erasing staging annotations does preserve semantics for certain
classes of staged programs. This raises a natural question by
\citet{DBLP:conf/esop/InoueT12, DBLP:journals/jfp/InoueT16}: ``\emph{when do we
know that erasing annotations preserves semantics}''?
They partially answered this question by developing applicative bisimulation techniques
for proving the equivalence of \emph{specific} staged programs and their
unstaged counterparts.  However, these techniques require substantial manual
proof effort and remain program-specific: they do not yield a general
erasure-soundness guarantee for the language as a whole.
This leaves a fundamental open question: can we obtain stronger semantic
guarantees other than well-typedness of generated code for a call-by-value
staging language?

To shed light on this question, let us turn to practical staging languages and
frameworks.  They often adopt \emph{let-insertion}
\cite{DBLP:conf/pldi/FlanaganSDF93} as a remedy, which has long been used in
partial evaluation for effectful programs \cite{DBLP:journals/jfp/KameyamaKS11,
DBLP:conf/popl/Danvy96, DBLP:conf/tacs/LawallT97, danvy2006analytical, dussart1997partial}.
The idea is to introduce let-binding at the right places to force the evaluation
and intended effects of a code fragment to happen, and then refer to the
let-bound variable in the rest of the code.
In this way, let-insertion preserves both effects \emph{and} their relative
evaluation order (see \Cref{sec:motivation} for illustrative examples).

One may then ask whether an \emph{automatic let-insertion} staging semantics
makes staged programs easier to reason about. More specifically, does it
guarantee semantics preservation of staging?
After all, let-insertion is widely used in practical staging systems precisely
because it is believed to help preserve the intended semantics.
Answering it, however, requires a formal account relating three artifacts: the
staged program, the code it generates, and its unstaged counterpart. Despite
substantial research in MSP on typing, hygiene, and intensional analysis,
this relational property has received far less attention.
As a result, whether let-insertion provides stronger reasoning principles
or semantics preservation remains largely folklore.

\subsubsection*{\textbf{This Work}}
To gain a rigorous understanding of the semantic guarantees provided by
let-insertion, this paper develops novel typed staging calculi featuring
\emph{automatic let-insertion} and investigates their metatheory.
We summarize the main results below; \Cref{sec:keyideas} provides an
overview for key ideas in our calculi, while the remainder of the paper
develops the formal theory.

Following \citet{DBLP:conf/esop/InoueT12, DBLP:journals/jfp/InoueT16}, we begin
with a pure setting without side effects, while still allowing divergence. In
this setting, we introduce \lang, a call-by-value, statically typed, two-stage
$\lambda$-calculus with general recursion and automatic let-insertion (\Cref{sec:core}).
Its reduction semantics follows \citet{DBLP:journals/pacmpl/AminR18}'s
untyped $\lambda_{\uparrow\downarrow}$-calculus, which already captures let-insertion
operationally (\Cref{sec:core:dynamics}).
We develop a novel type-and-effect system for automatic let-insertion
(\Cref{sec:core:statics}), where
the key idea is to treat code generation as a control effect in the first stage
and to distinguish complete code values from intermediate code fragments.

We prove syntactic type soundness (\Cref{sec:syntactic-soundness}),
define the erasure operation, and show that erasure preserves syntactic
typing (\Cref{sec:erasure}), which is a prerequisite for semantics preservation.
We further prove the semantics-preservation theorem for staging and erasure (\Cref{sec:lr}).
The central technical tool is a binary Kripke logical relation for the two-stage
calculus: we show that if a well-typed two-stage program $t_1$
evaluates to code $t_2$, then $t_2$ is contextually equivalent to the erased
term $|t_1|$, where $|\cdot|$ removes all staging annotations.
Thus, in \lang, erasure is sound for \emph{all} well-typed programs.
This directly answers the question posed by \citet{DBLP:conf/esop/InoueT12,
DBLP:journals/jfp/InoueT16}: \emph{ erasing annotations in \lang always
preserves semantics}.

Next, we introduce ML-style mutable references to study the interaction between
let-insertion and store effects (\Cref{sec:mutref}).
Unrestricted references reveal an important boundary of erasure soundness:
staging may evaluate under second-stage binders or conditionals, producing
first-stage store effects that erasure does not preserve
(\Cref{sec:when-break}).
We then recover erasure soundness for a broad class of programs by restricting
references to the second stage. We formalize this approach in \langref, a modest
extension of \lang (\Cref{sec:langref-syntax}), and prove both type soundness
and erasure soundness (\Cref{sec:langref-sem}).  This result is practically
important because generating and manipulating effectful code is a central use
case for staging.

At the same time, our study proposes
a challenging open question: how to support more expressive first-stage store effects while
retaining erasure soundness. We discuss the obstacles and possible directions (\Cref{sec:extension}),
and hope this work will serve as a foundation for future progress on
semantics-preserving staging.
Finally, all calculi and their metatheory results are mechanized
(\Cref{sec:mechanization}) in Lean~4~\cite{DBLP:conf/cade/Moura021}.

\subsubsection*{\textbf{Contributions and Organization}}
The main contributions of this paper are threefold:

\begin{itemize}[leftmargin=1.7em]

\item To the best of our knowledge, no prior work has developed a typed,
semantics-preserving staging calculus with a mechanized erasure soundness proof.
We close this gap and answer the open question raised by
\citet{DBLP:conf/esop/InoueT12, DBLP:journals/jfp/InoueT16} by developing the
first typed CBV multi-stage calculi \lang and \langref, with mechanized erasure
soundness proofs.
Our contributions allow programmers to introduce staging annotations for performance
without worrying that they will silently alter the program's meaning.

\item We develop a novel type-and-effect system for automatic let-insertion.
The key idea is to treat code generation as a control effect, while
distinguishing complete code values from intermediate code fragments.
The $\langref$-calculus presents a novel design combining first-stage
control effects and second-stage store effects, ensuring sound
manipulation of effectful code.

\item We apply binary Kripke logical relations to relate staged and unstaged programs
and to prove semantics preservation of staging. To the best of our knowledge,
this is one of the first uses of logical relations to to reason about staging semantics
and relate generated code to its erased counterpart,
and we expect the technique to be of independent interest as a
reusable foundation for future work.
\end{itemize}

\noindent
The rest of the paper is organized as follows.

\begin{itemize}[leftmargin=1.7em]
  \item In \Cref{sec:motivation}, we
    motivate the semantics preservation issue by examples
    that manipulate effectful code, discuss existing remedies,
    and introduce the key ideas of \lang/\langref.
  \item In \Cref{sec:core}, we present the syntax, reduction semantics, and type-and-effect system of \lang.
    In \Cref{sec:syntactic-soundness}, we prove the syntactic type soundness of \lang via
    progress and preservation.
  \item In \Cref{sec:erasure}, we introduce the erasure operation applied to
    terms, types, and environments, and prove the syntactic erasure soundness
    theorem. In \Cref{sec:lr}, we introduce the notion of contextual equivalence,
    develop a step-indexed Kripke logical relation for the erased subset language
    of \lang, and prove the semantic soundness of erasure via the logical relation.
  \item In \Cref{sec:mutref}, we discuss
    the challenges in supporting general store effects while retaining erasure
    soundness. Based on the discussion, we extend \lang to \langref with
    second-stage only mutable references,
    and prove syntactic type soundness and erasure soundness for it.
  \item In \Cref{sec:extension}, we discuss possible solutions to support more
    expressive first-stage effects while retaining erasure soundness.  We also
    discuss other possible extensions, such as supporting more stages and
    higher-order store.
  \item In \Cref{sec:mechanization}, we discuss the mechanization in Lean 4.
        In \Cref{sec:related}, we discuss related work.
\end{itemize}

\section{Motivation and Overview} \label{sec:motivation}

In this section, we present a motivating example to illustrate the issue of
semantics preservation in staging and existing solutions, and then
explain the key ideas and main results of our work.

\subsection{Motivating Example: Generative Imperative Power}

The canonical motivating example for MSP is specializing the power function --- \ie,
computing $x^n$ when the exponent $n$ is known at code generation time.
The power function can be implemented in a purely functional style, but
to discuss the subtle interaction between staging and effects, we consider an
imperative version that uses a mutable reference \code{res} to accumulate the result:
\begin{lstlisting}[language=OCaml]
  let rec power' (n: int) (x: int) (res: int ref) : int =
    if n = 0 then !res
    else (res := x * !res; power' (n - 1) x res)
\end{lstlisting}
To call \code{power'}, the client code allocates a reference initialized to
\code{1} and passes it as an argument:
\begin{lstlisting}[language=OCaml]
  let power n = fun x -> power' n x (ref 1)
\end{lstlisting}

\subsubsection*{\textbf{Staged Imperative Power via Quasi-Quotation}}
Now, given a known value for \code{n}, a careless programmer may attempt to
specialize \code{power'} by simply adding stage annotations.
In this example, we use MetaML-style syntax for quotation and splicing.
Since \code{x} and \code{res} are only known at run-time, they
are marked as \code{code} types, while \code{n} is a plain \code{int};
the return type is also \code{int code}.
\begin{lstlisting}[language=OCaml]
  let rec power' (n: int) (x: int code) (res: (int ref) code) =
    if n = 0 then .< !(.~res) >.
    else .<.~res := .~x * !(.~res); .~(power' (n - 1) x res)>.
\end{lstlisting}
The body has been adapted using proper quotations and splices.
The two branches of the \code{if} expression are both quoted code fragments.
When referring to \code{x} and \code{res} immediately inside quotations, we
splice (\eg, \code{.~x}) them as they are already code values.
To drive the specialization, we use \code{run} to execute the generated code at runtime.
In the code argument for \code{run}, the function provides parameter \code{x}
that can be used as a placeholder for the argument to be passed at runtime:
\begin{lstlisting}[language=OCaml]
  let power n = run .<fun x -> .~(power' n .<x>. .<ref 1>.)>.
\end{lstlisting}
We also add quotations around the allocation, since
the result should be produced in the second stage, and
we are in a splicing context to call \code{power'}.

However, despite the term ``annotations'' suggesting that they are orthogonal to
program behavior, staging here unrolls the recursion and generates code that
duplicates all allocations:
\begin{lstlisting}[language=OCaml]
  fun x -> (* example: specialized for n = 2 *)
    (ref 1) := x * !(ref 1);
    (ref 1) := x * !(ref 1);
    !(ref 1)
\end{lstlisting}
The generated code degenerates to a constant function that always returns 1,
which is clearly not the intended power function!
The problem here is that the quoted allocation \code{res} binds in the first stage,
and each occurrence of \code{res} simply substitutes it into the larger template,
causing duplication.
When manipulating effectful code during staging, such duplication may violate
the intended semantics of the generated code.

To generate the intended code, the programmer must manually insert
a quoted let-binding to name the allocation, thus allocating only once:
\begin{lstlisting}[mathescape=false, language=OCaml]
  let power n = run .<fun x -> let r = ref 1 in .~(power' n .<x>. .<r>.)>.
\end{lstlisting}
MetaOCaml \cite{DBLP:journals/scp/Kiselyov26} also provides an auxiliary primitive \code{genlet} to insert
let-bindings for effectful code, which can be used as follows, avoiding an explicit \code{let}-expression:
\begin{lstlisting}[mathescape=false, language=OCaml]
  let power n = run .<fun x -> .~(power' n .<x>. (genlet .<ref 1>.))>.
\end{lstlisting}
In either case, when refactoring an unstaged program into a staged one,
the programmer must be peculiarly aware of whether the meta-program manipulates
effectful code, and must manually indicate let-bindings to avoid duplication
and unintended changes in behavior.

\subsubsection*{\textbf{Staging with Automatic Let-Insertion}}
We have seen that simply adding staging annotations is not enough to
correctly stage the \emph{imperative} power function:
without proper let-insertion, syntactic quotation does not in general
preserve semantics.
This is precisely why practical staging frameworks such as LMS, Squid, and BuildIt
support \emph{automatic let-insertion}, relieving programmers from having
to insert let-bindings by hand.

To illustrate automatic let-insertion, we now switch from the quotation-based
staging syntax to \emph{combinator-based} syntax.
Instead of building code with quotation and splicing,
we use combinators that construct code values directly.
This style goes back to two-level functional languages \cite{DBLP:books/daglib/0071307}
and partial evaluation \cite{DBLP:books/daglib/0072559}.
In practice, LMS elaborates surface syntax into such code-generating combinators via
Scala's overloading mechanisms \cite{DBLP:conf/birthday/Rompf16},
and MetaOCaml translates quotations/splices into a
combinator-based representation \cite{DBLP:journals/scp/Kiselyov26} similar to ours.

The following example shows a semantics-preserving staged version of the
imperative power function in \langref\ (\Cref{sec:mutref}).
Rather than annotating with quotation and splicing,
the programmer marks (infix) \emph{operations} with the stage annotation $\mathbb{2}$.
For instance, dereference, assignment, and multiplication on \code{res} are
all marked with $\mathbb{2}$, indicating that they should occur at the second stage.
Variables themselves require no splicing or annotation; they simply refer to code values.
\begin{lstlisting}[language=OCaml]
  let rec power' (n: int) (x: int code) (res: (int ref) code) : int code =
    if n = 0 then !$^\mathbb{2}$res
    else (res :=$^\mathbb{2}$ x *$^\mathbb{2}$ !$^\mathbb{2}$res; power' (n - 1) x res)
\end{lstlisting}
To drive specialization,
we residualize the allocation,
use \code{lift} to promote first-stage values to the second stage,
and apply \code{run} to obtain the specialized function:
\begin{lstlisting}[language=OCaml]
  let power n = run (lift (fun x -> power' n x (ref$^\mathbb{2}$ (lift 1))))
\end{lstlisting}
Of course, the combinator-based syntax alone does not guarantee let-insertion.
What matters is the dynamic semantics of \langref, which produces the desired residual program that runs faster:
\begin{lstlisting}[language=OCaml]
  fun x -> let res = ref 1 in (* example: specialized for n = 2 *)
    res := x * !res; res := x * !res; !res
\end{lstlisting}
Crucially, the generated code allocates \code{res} only once, and all subsequent
dereferences and assignments refer to that same location.
The programmer does not need to insert any let-bindings manually, yet the generated
code can still serve as a drop-in replacement for computing $x^2$.

\subsection{Key Ideas and Main Results} \label{sec:keyideas}

In the example above, the let-inserted staged \code{power} clearly generates code
semantically equivalent to the reference implementation, namely the unstaged \code{power}
function.
However, formally establishing this property is far from trivial, especially in the
presence of the subtle interactions between staging annotations and effects.
Our approach is to connect staged and unstaged programs via
\emph{erasure}, and connect generated code and unstaged programs via
\emph{contextual equivalence}.
\Cref{fig:proof} illustrates the high-level proof structure of our approach.
Below, we present the key ideas behind our calculi
and give an overview of our results.

\subsubsection*{\textbf{Syntax and Dynamic Semantics}}

The \lang- and \langref-calculi share the same foundation of syntax and semantics, with
the latter extending the former with mutable references.
We adopt the dynamic semantics of \citet{DBLP:journals/pacmpl/AminR18}'s untyped
$\lambda_{\uparrow\downarrow}$-calculus, which formulates let-insertion as a
reduction semantics and captures operational aspects found in practical
staging frameworks (such as LMS
\cite{DBLP:journals/cacm/RompfO12,DBLP:conf/gpce/RompfO10,DBLP:conf/birthday/Rompf16})
and partial evaluation for sound specialization with effects
\cite{DBLP:journals/jfp/KameyamaKS11,DBLP:conf/popl/Danvy96,
DBLP:conf/tacs/LawallT97,danvy2006analytical,dussart1997partial}.
With let-insertion, generated code is organized as a sequence of let-bindings,
similar to A-normal form (ANF)
\cite{DBLP:conf/pldi/FlanaganSDF93} and monadic normal form (MNF)
\cite{DBLP:conf/popl/HatcliffD94}.

Like MetaML, our calculi provide a \code{run} operation for executing
generated code at runtime and a \code{lift} operation for promoting first-stage
values to second-stage values.
However, unlike MetaML's \code{lift}, which is restricted to ground values,
\code{lift} in our calculi also applies to $\lambda$-terms, where it acts as a
two-level $\eta$-expansion and evaluates the function body \cite{DBLP:journals/pacmpl/AminR18,DBLP:journals/toplas/DanvyP96}.
For example, the following code uses \code{lift} to promote a first-stage
function of \lstinline[language=OCaml]|int code -> int code| to the second stage:
\begin{lstlisting}[language=OCaml]
  let f1 (x: int code) : int code = x +$^\mathbb{2}$ (lift 1) +$^\mathbb{2}$ (lift 2) in
  lift f1
\end{lstlisting}
The resulting generated code has type \lstinline[language=OCaml]|(int -> int) code|.
Its body sequentializes the evaluation order through a sequence of proper let-expressions:
\begin{lstlisting}[language=OCaml]
  fun x -> let x0 = x + 1 in let x1 = x0 + 2 in x1$.$
\end{lstlisting}

\subsubsection*{\textbf{Type-and-Effect System}}

Although automatic let-insertion has been formalized in prior work~\cite{DBLP:journals/pacmpl/AminR18},
a type system that guarantees type soundness is new in this work.
Keeping erasure soundness in mind, the challenge is that code cannot be
treated as ordinary pure values: they cannot be duplicated, reordered, or discarded arbitrarily.
A simply typed system cannot rule out programs that violate these constraints.
This issue is particularly acute for effectful second-stage code, whose
let-insertion points must precisely preserve effects and evaluation
order.
For example, consider the following program, which attempts to construct second-stage
\code{print} code yet discards it:
\begin{lstlisting}[language=OCaml]
  let x = print$^\mathbb{2}$ "Hello" in 42
\end{lstlisting}
Since \code{x} is unused, one might expect the first stage to evaluate to
\code{42}, thereby discarding the generated \code{print}.
But admitting this program as well-typed would violate semantics preservation,
since the erased program should still print \code{"Hello"}.

To rule out such programs, we develop lightweight type-and-effect systems for
\lang and \langref that treat let-inserting code generation
uniformly as a control effect rather than as pure value construction.
The control effect identifies a delimited reification boundary around the
focused expression and lifts that expression to a let-binding within that
boundary.

Accordingly, the type system distinguishes intermediate code fragments
($\tyfrag{\tau}$) from complete code values ($\tyrep{\tau}$).
Code fragments of type $\tyfrag{\tau}$ are introduced by let-insertion.
Once named by code variables, they are pure and can be used freely.
Complete code values $\tyrep{\tau}$, in contrast, represent fully generated
code that may contain arbitrary second-stage effects such as divergence,
and thus cannot be duplicated or discarded arbitrarily.
The type system tracks pending reification as a control effect through an
effect flag, ensuring that fragments are properly let-inserted before being
used.

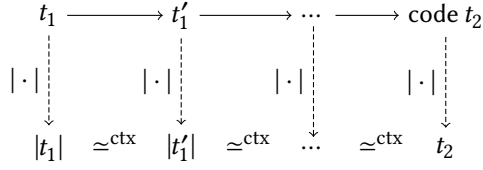
\begin{figure}
\begin{tikzpicture}[node distance=1.75cm, auto]

\node (a) {$t_1$};
\node (b) [right of=a] {$t_1'$};
\node (e) [right of=b] {$\dots$};
\node (f) [right of=e] {$\tmcode{t_2}$};

\node (a') [below of=a] {$|t_1|$};
\node (b') [below of=b] {$|t_1'|$};
\node (e') [below of=e] {$\dots$};
\node (f') [below of=f] {$t_2$};

\draw[->] (a) -- (b);
\draw[->] (b) -- (e);
\draw[->] (e) -- (f);

\draw[->, dashed, dash pattern=on 2pt off 1pt] (a) -- (a') node[midway, left] {$|\cdot|$};
\draw[->, dashed, dash pattern=on 2pt off 1pt] (b) -- (b') node[midway, left] {$|\cdot|$};
\draw[->, dashed, dash pattern=on 2pt off 1pt] (e) -- (e') node[midway, left] {$|\cdot|$};
\draw[->, dashed, dash pattern=on 2pt off 1pt] (f) -- (f') node[midway, left] {$|\cdot|$};

\draw[-, draw=none] (a') -- (b') node[midway, yshift=-6pt] {$\simeq^{\text{ctx}}$};
\draw[-, draw=none] (b') -- (e') node[midway, yshift=-6pt] {$\simeq^{\text{ctx}}$};
\draw[-, draw=none] (e') -- (f') node[midway, yshift=-6pt] {$\simeq^{\text{ctx}}$};

\end{tikzpicture}
\vspace{-1em}
\caption{Illustration of semantics preservation and our proof structure.}
\label{fig:proof}
\vspace{-1em}
\end{figure}

\subsubsection*{\textbf{Metatheory}}

By combining a first-stage effect system with a type-level distinction between
complete code values and intermediate code fragments, we obtain a simple and
robust foundation for proving syntactic type soundness, \ie, progress and
preservation, which further ensures that the generated program
is also well-typed.

Our main technical contribution is the semantics-preservation theorem relating
staging annotations to their inverse operation, erasure.
To show that, instead of reasoning about the staged program itself,
we reason about the result of such a staged program (\ie, the generated code) and its
equivalence with the erased program via binary step-indexed Kripke logical relations.
Informally, the theorem (\Cref{sec:lr}) states that if a well-typed
two-stage program $t_1$ generates code $t_2$, then $t_2$ is contextually
equivalent to the erased program $|t_1|$, where $|\cdot|$ removes all staging
annotations.
As illustrated in \Cref{fig:proof}, we establish a stronger stepwise property:
every reduction step preserves contextual equivalence between erased terms,
\ie, whenever $t \rightarrow t'$, we have $|t| \simeq^{\text{ctx}} |t'|$.
It then follows that
$|t_1| \simeq^{\text{ctx}} t_2$ where
$t_1 \to^* \mathsf{code}~t_2$,
since the erasure of $\mathsf{code}~t_2$ is just $t_2$.

\subsubsection*{\textbf{Extending with Store Effects}}

The base calculus $\lang$ already admits divergence, and we show that
let-insertion preserves semantics between staged programs and their erasures,
yielding strong call-by-value reasoning principles for multi-stage programming.
Our extension with mutable references further clarifies the scope of this
guarantee: contrary to common intuition, let-insertion alone does not ensure
semantic preservation in the presence of arbitrary effects during staging.

To study this interaction, we extend $\lang$ with mutable references, yielding
\langref. In \Cref{sec:mutref}, we characterize how first-stage store effects
can violate semantic preservation and establish a useful sufficient
condition: restricting store effects to the second stage. This restriction
supports a common use case, \ie, soundly generating and manipulating effectful
code, and lays the foundation for future type-and-effect systems that distinguish
unsafe first-stage effects from admissible ones. By adapting both the reduction
semantics and logical relation, we prove semantic preservation for \langref.
We further discuss possible paths toward a more expressive system
(\Cref{sec:extension}), drawing on recent type systems for tracking
capture/reachability and establishing separation
\cite{DBLP:journals/toplas/BoruchGruszeckiOLLB23,
DBLP:journals/pacmpl/WeiBJBR24,
DBLP:conf/esop/MarshallVO22,
DBLP:journals/pacmpl/BaoWBJHR21,
DBLP:journals/pacmpl/XuBO24}.

\section{\texorpdfstring{\lang: }{}A Semantics-Preserving Two-Stage Calculus with Recursion} \label{sec:core}

We now formally present the base \lang-calculus that enjoys erasure soundness
via automatic let-insertion.
Following \citet{DBLP:conf/esop/InoueT12, DBLP:journals/jfp/InoueT16},
we consider a pure setting without mutable references but allow divergence.
\lang is a statically typed, call-by-value, two-stage $\lambda$-calculus with
general recursion and automatic let-insertion.
It specializes the $\lambda_{\uparrow\downarrow}$-calculus's reduction semantics
\cite{DBLP:journals/pacmpl/AminR18} to two stages and equips it with a
type-and-effect system.
Building on the $\lambda_{\uparrow\downarrow}$'s account of let-insertion, our
type system ensures both syntactic type soundness and semantic preservation
under erasure.

We first present the surface syntax of \lang, then its dynamic semantics with
administrative and runtime constructs, and finally its type-and-effect system.

\subsection{Surface Syntax}  \label{sec:core:syntax}

\begin{figure}
\[
\begin{array}{rcll}
& n & \in & \mathbb{N}     \hspace{4em}  x \hspace{0.7em} \in \hspace{0.7em}  \textsf{Vars}
  \hspace{4em} \textsf{Stages}  \hspace{0.7em}   s  \hspace{0.7em}   \in \hspace{0.7em} \{ \mathbb{1}, \mathbb{2} \} \\
\textsf{Terms}  & t & ::= &
  \texttt{()}
  \mid n
  \mid x
  \mid \lambda x. t
  \mid \Keywd{let} ~x = t_1~\Keywd{in}~t_2
  \mid \Keywd{lift} ~ t
  \mid \Keywd{run} ~ t \\
  & & \hspace{0.5em} \mid &
  \Keywd{app}^s ~ t_1 ~ t_2
  \mid \Keywd{fix}^s ~ t
  \mid \Keywd{ifz}^s ~ t_1 ~ t_2 ~ t_3
  \mid t_1 \oplus^s t_2
\end{array}
\]
\vspace{-1em}
\caption{Surface syntax of \lang.}
\label{fig:syntax_pure}
\vspace{-1em}
\end{figure}

\Cref{fig:syntax_pure} presents the surface syntax of \lang.
The syntax follows the two-level $\lambda$-calculus \cite{DBLP:books/daglib/0071307};
we adopt Curry-style syntax and defer the type system to \Cref{sec:core:statics}.
\lang is designed to manipulate and generate second-stage code.
Stage superscripts $s \in \{\mathbb{1}, \mathbb{2}\}$ annotate function application
($\mathsf{app}^s$), arithmetic operations ($\oplus^s$), fixed-point
($\mathsf{fix}^s$), and conditionals ($\mathsf{ifz}^s$).
Unlike the $\lambda_{\uparrow\downarrow}$-calculus \cite{DBLP:journals/pacmpl/AminR18}
that uses self-references of $\lambda$-terms for recursion, we use a separate
fixed-point operator (the Z combinator) that is compatible with call-by-value evaluation.

Let-bindings in the surface language are stage-agnostic: they are evaluated in the first stage, yet the bound
term and body can be first- or second-stage.
Simple expressions (unit, numbers, variables, $\lambda$-abstractions) are
annotation-free and can be lifted.
Given a first-stage term $t$, \tmlift{t} evaluates $t$ and produces a code fragment
representing its value.
Conversely, \tmrun{t} evaluates a
code value \tmcode{t} and produces a first-stage value (similar to \code{eval} in Lisp).
Together with \tmlift{t} and \tmrun{t}, they support runtime code generation and
evaluation.

\subsection{Dynamic Semantics} \label{sec:core:dynamics}

\begin{figure}
    \[
    \begin{array}{rcll}
    \textsf{Terms}  & t & ::= & \dots \mid g \\
    \textsf{Administrative Terms} & g & ::= &
      \Keywd{code} ~ t
      \mid \Keywd{reflect} ~ t
      \mid \Keywd{let_c}~x = t_1~\Keywd{in}~t_2
      \mid \lambda_\mathsf{c} x. t \\
    \textsf{Values} & v & ::= & n
      \mid \texttt{()}
      \mid \lambda x. t
      \mid \Keywd{code} ~ t \\
    \textsf{Pure Frames} & B & ::= &
      \mathsf{app}^s ~ \square ~ t
      \mid \mathsf{app}^s ~ v ~ \square
      \mid \square ~ \oplus^s ~ t
      \mid v ~ \oplus^s ~ \square
      \mid \mathsf{fix}^s ~ \square \\
      & & \hspace{0.5em} \mid &
      \mathsf{ifz}^s ~ \square ~ t_2 ~ t_3
      \mid \mathsf{lift} ~ \square
      \mid \mathsf{let} ~ x = \square ~ \Keywd{in} ~ t
      \\
    \textsf{Pure Contexts} & E & ::= & \square \mid B \circ E \\
    \textsf{Reification Frames} & R & ::= &
      \lambda_\mathsf{c} x. \square
      \mid \mathsf{let_c} ~ x = t ~ \Keywd{in} ~ \square
      \mid \mathsf{run} ~ \square
      \mid \mathsf{ifz}^\mathbb{2} ~ v ~ \square ~ t
      \mid \mathsf{ifz}^\mathbb{2} ~ v_1 ~ v_2 ~ \square \\
    \textsf{Reification Contexts} & P & ::= & \square \mid Q \\
                                  & Q & ::= & R \mid B \circ Q \mid R \circ Q \\
    \textsf{Evaluation Contexts} & M & ::= & \square \mid B \circ M \mid R \circ M \\
    \end{array}
    \]

    \judgement{Reduction}{ \BOX{t \headto t'} \BOX{t \to t'}}
    $$
    \begin{array}{lrclr}
    \steprule{st-let} & \tmlet{x}{v}{t} & \headto & t[v/x]                                     & \\
    \steprule{st-app$^\mathbb{1}$} & \tmappSt{(\lambda x. t)}{v} & \headto & t[v/x]                         & \\
    \steprule{st-app$^\mathbb{2}$} & \tmappDy{(\tmcode{t_1})}{(\tmcode{t_2})} & \headto & \tmreflect{(\tmappSt{t_1}{t_2})} & \\
    \steprule{st-op$^\mathbb{1}$} & n_1 \oplus^\mathbb{1} n_2 & \headto & n_3  \hspace{7em} \text{where}~ n_3 = \delta(\oplus, n_1, n_2) & \\
    \steprule{st-op$^\mathbb{2}$} & (\tmcode{t_1}) \oplus^\mathbb{2} (\tmcode{t_2}) & \headto & \tmreflect{( t_1 \oplus^\mathbb{1} t_2 )} & \\
    \steprule{st-nat$^\uparrow$} & \tmlift{n} & \headto & \tmreflect{n} & \\
    \steprule{st-unit$^\uparrow$} & \tmlift{\texttt{()}} & \headto & \tmreflect{\texttt{()}} & \\
    \steprule{st-lam$^\uparrow$} & \tmlift{(\lambda x.t)} & \headto & \tmlamc{x}{t[\tmcode{x}/x]} & \\
    \steprule{st-lam$_\mathsf{c}$} & \tmlamc{x}{(\tmcode{t})} & \headto & \tmreflect{(\lambda x.t)} & \\
    \steprule{st-let$_\mathsf{c}$} & \tmletc{x}{t_1}{(\tmcode{t_2})} & \headto & \tmcode{(\tmlet{x}{t_1}{t_2})}   & \\
    \steprule{st-run} & \tmrun{(\tmcode{t})} & \headto & t                                     & \\
    \steprule{st-fix$^\mathbb{1}$} & \tmfixSt{v} & \headto & \lambda x. \tmappSt{ (\tmappSt{v}{(\tmfixSt{v})}) }{x} & \\
    \steprule{st-fix$^\mathbb{2}$} & \tmfixDy{(\tmcode{t})} & \headto & \tmreflect{(\tmfixSt{t})} & \\
    \steprule{st-ifz$^\mathbb{1}$-0} & \tmifzSt{0}{t_1}{t_2} & \headto & t_1                                   & \\
    \steprule{st-ifz$^\mathbb{1}$-n} & \tmifzSt{n}{t_1}{t_2} & \headto & t_2 \hfill \text{where}~ n \neq 0 & \\
    \steprule{st-ifz$^\mathbb{2}$} & \tmifzDy{(\tmcode{t_1})}{(\tmcode{t_2})}{(\tmcode{t_3})} & \headto & \tmreflect{(\tmifzSt{t_1}{t_2}{t_3})} & \\
    \end{array}
    $$
    \begin{minipage}[t]{.27\linewidth}
      \infrule[st-pure]{
        t \headto t'
      }{
        M[t] \to M[t']
      }
    \end{minipage}
    \begin{minipage}[t]{.63\linewidth}
      \infrule[st-reflect]{
        x~\text{is~fresh}
      }{
        P[E[\tmreflect{t}]] \to P[\tmletc{x}{t}{E[\tmcode{x}]}]
      }
    \end{minipage}
\caption{Additional syntax and reduction semantics of \lang.}
\vspace{-1em}
\label{fig:reduction_pure}
\end{figure}

The key aspect of \lang's dynamic semantics is let-insertion,
formalized as reduction steps following \citet{DBLP:journals/pacmpl/AminR18}.
\Cref{fig:reduction_pure} defines the additional syntactic categories
required by this semantics, including the administrative terms, values, and evaluation
contexts.
Administrative terms $g$ occur only at runtime and comprise code fragments
$\Keywd{code}~t$, reflections $\Keywd{reflect}~t$, code let-bindings
$\Keywd{let_c}~x=t_1~\Keywd{in}~t_2$, and code functions
$\lambda_\mathsf{c} x.\, t$.
Values are those of the standard $\lambda$-calculus, extended with code values
of the form $\Keywd{code}~t$.

\subsubsection*{\textbf{Two-Stage Reduction Contexts}}
Evaluation in \lang proceeds in two stages under a small-step reduction
relation.
First, a term $t$ reduces to a value $v$, written $t \to^* v$.
When $v = \tmcode{t'}$, the second stage evaluates the generated term $t'$ to
produce the final value.
As \lang is a two-stage language, $t'$ contains no remaining staging
constructs, \ie, no \textsf{lift}, \textsf{run}, or $\mathbb{2}$-annotations.

\Cref{fig:reduction_pure} defines the pure
context $E$ for standard CBV evaluation and the reification context $P$ for let-insertion,
under which the reduction rules apply.
Pure contexts $E$ is either empty (\ie, a hole), or a sequence of pure frames
$B$, which are standard evaluation frames (\ie, single terms with a hole)
for left-to-right, call-by-value evaluation.
We call them ``pure'' because they do not constrain the scope of let-insertion.
Context composition $B \circ E$ denotes
plugging $E$ into the hole of $B$, i.e., $B[E]$, which is still a context.

Reification contexts $P$ are either empty or $Q$-contexts.
Context $Q$ is built from reification frames $R$ and pure frames $B$:
the innermost frame in $Q$ must be an $R$-frame, which identifies the
let-insertion boundary when it contains a reflection $\tmreflect{t}$.
$R$-frames allow evaluation under second-stage binders (\eg, in a $\lambda_\mathsf{c}$- or
$\mathsf{let_c}$-body), inside branches of second-stage conditionals, or under
$\mathsf{run}$.
These $P$- and $E$-contexts mark the boundary where let-insertion can occur; the
reduction rules below make this precise.
Full evaluation contexts $M$ allow any interleaving of $B$ and $R$.

\subsubsection*{\textbf{Pure Reduction}}
In \Cref{fig:reduction_pure}, $t \headto t'$ denotes the single-step
reduction under an evaluation context $M$ (\textsc{st-pure}).
Rules annotated with $\mathbb{1}$ are standard call-by-value reduction;
$\delta$ denotes primitive arithmetic in \textsc{st-op}$^\mathbb{1}$.
The first-stage fixed-point operator (\textsc{st-fix$^\mathbb{1}$})
$\eta$-expands the function value $v$ with the fixed function itself
and the ordinary argument, following the Z combinator.

Reduction rules annotated with $\mathbb{2}$ ``reflect'' the corresponding first-stage
term when all subterms have evaluated to code values.
For example, \textsc{st-app$^\mathbb{2}$} steps to $\tmreflect{(\tmappSt{t_1}{t_2})}$,
where $t_1$ and $t_2$ are the function and argument code fragments, respectively.

Lifting a value likewise steps to reflecting. For a $\lambda$-value,
lifting replaces the bound variable $x$ with the neutral code variable
$\tmcode{x}$ in the body $t$ (\textsc{st-lam$^\uparrow$}), yielding $\tmlamc{x}{t[\tmcode{x}/x]}$.
Any latent first-stage computation in the resulting body may then reduce further.
This rule follows the normalization-by-evaluation
\cite{DBLP:conf/nada/BergerES98, DBLP:conf/lics/BergerS91} and type-directed
partial-evaluation \cite{DBLP:conf/popl/Danvy96} technique of normalizing a
function by applying it to a neutral variable. It also models the lifting
mechanism used in staging frameworks such as LMS
\cite{DBLP:journals/cacm/RompfO12, DBLP:conf/scala/Rompf16}.

\subsubsection*{\textbf{Reduction with Let-Insertion}}
The let-insertion rule (\textsc{st-reflect}) is central to \lang's dynamics.
A term $\tmreflect{t}$ marks a let-insertion site but is not itself a redex.
When it appears within a reification context $P$ and a pure context $E$,
the rule inserts a $\mathsf{let_c}$-binding $x = t$ where $x$ is fresh, and replaces
$\tmreflect{t}$ with $\tmcode{x}$ in context $E$.
Here, $P$ is the largest context whose innermost frame is $R$ and may contain
$E$. Because $t$ cannot depend on bindings introduced by the inner $E$, hoisting the
binding into $P$ is sound. The resulting $\mathsf{let_c}$-binding represents an
intermediate code-generation step:
the right-hand side has been reflected/reified (thus represented
as a first-stage term), while the body may still contain unevaluated code.
Once the body becomes a code value, rule (\textsc{st-let$\mathsf{c}$}) reduces
the expression to code containing an ordinary let-binding.

\subsubsection*{\textbf{Determinism}}
A term decomposes uniquely as either $M[t_1]$ or $P[E[\tmreflect{t_2}]]$,
and the $P$/$E$-context decomposition is also unique,
so exactly one of \textsc{st-pure} and \textsc{st-reflect} applies.
We later use determinism to prove transitivity of contextual equivalence
(\Cref{theorem:transitivity}).

\begin{lemma}[Deterministic Decomposition]
If $P[E[\tmreflect{t}]] = P'[E'[\tmreflect{t'}]]$, then $P = P'$, $E = E'$, and $t = t'$.
\end{lemma}

\begin{theorem}[Determinism]
\label{theorem:deterministic}
If\, $t_1 \to t_2$ and $t_1 \to t_3$, then $t_2 = t_3$.
\end{theorem}

\subsection{Static Semantics} \label{sec:core:statics}

\begin{figure}
\[
\begin{array}{rcll}
\textsf{Reification Effects} & \epsilon & ::= & \bot \mid \top \\
\textsf{Types} & \tau & ::= &
  \Keywd{unit}
  \mid \Keywd{nat}
  \mid \tau_1 \to^{\epsilon} \tau_2
  \mid \tyrep{\tau}
  \mid \tyfrag{\tau} \\
\textsf{Typing Environments} & \Gamma & ::= &
  \varnothing
  \mid \Gamma, x^s : \tau
\end{array}
\]
\vspace{-1em}
\caption{Syntax for \lang's static semantics.}
\vspace{-1em}
\label{fig:static-syntax-pure}
\end{figure}
\begin{figure}
\judgement{Type and Binding-time Well-formedness}{ \BOX{\WF^s ~ \tau}}\\[1ex]
\begin{minipage}[t]{.39\linewidth}
  \infrule[wf-arrow]{
    \WF^{s} ~ \tau_1 \quad
    \WF^{s} ~ \tau_2 \quad 
    s = \mathbb{2} \implies \epsilon = \bot
  }
  {
    \WF^{s} ~ (\tau_1 \to^{\epsilon} \tau_2)
  }
\end{minipage}
\begin{minipage}[t]{.30\linewidth}
  \infrule[wf-base]{
    \tau = \Keywd{unit} \lor \tau = \Keywd{nat}
  }{
    \WF^s ~ \tau
  }
\end{minipage}
\begin{minipage}[t]{.25\linewidth}
  \infrule[wf-frag]{
    \WF^{\mathbb{2}} ~ \tau
  }{
    \WF^{\mathbb{1}} ~ (\tyfrag{\tau})
  }
\end{minipage}
\caption{Type and binding-time well-formedness.}
\label{fig:type-wf-pure}
\end{figure}

\begin{figure}[ht]
\judgement{Typing Judgments for Surface Terms}{ \BOX{\Gamma \vdash^s t : \tau \mid \epsilon} \BOX{\Gamma \vdash t : \tau \mid \epsilon}}\\[1ex]
\begin{minipage}[t]{.32\linewidth}
  \infrule[t-var]{
    x : \tau^s \in \Gamma
    \quad
    \WF^s ~ \tau
  }{
    \Gamma \vdash^s x : \tau \mid \bot
  }
\end{minipage}
\begin{minipage}[t]{.32\linewidth}
  \infrule[t-nat]{
  }{
    \Gamma \vdash^s n : \Keywd{nat} \mid \bot
  }
\end{minipage}
\begin{minipage}[t]{.32\linewidth}
  \infrule[t-unit]{
  }{
    \Gamma \vdash^s \texttt{()} : \Keywd{unit} \mid \bot
  }
  \vgap
\end{minipage}
\begin{minipage}[t]{.49\linewidth}
  \infrule[t-lam$^\uparrow$]{
    \Gamma \vdash^\mathbb{1} t : (\tyfrag{\tau_1} \to^{\epsilon_1} \tyfrag{\tau_2}) \mid \epsilon_2
  }{
    \Gamma \vdash^\mathbb{1} \Keywd{lift} ~ t : \tyfrag{(\tau_1 \to^{\bot} \tau_2)} \mid \top
  }
  \vgap
  \infrule[t-unit$^\uparrow$]{
    \Gamma \vdash^\mathbb{1} t : \Keywd{unit} \mid \epsilon
  }{
    \Gamma \vdash^\mathbb{1} \Keywd{lift} ~ t : \tyfrag{\Keywd{unit}} \mid \top
  }
  \vgap
  \infrule[t-app$^\mathbb{1}$]{
    \Gamma \vdash^s t_1 : (\tau_1 \to^{\epsilon} \tau_2) \mid \epsilon_1
    \\
    \Gamma \vdash^s t_2 : \tau_1 \mid \epsilon_2
  }{
    \Gamma \vdash^s \Keywd{app}^\mathbb{1} ~ t_1 ~ t_2 : \tau_2 \mid (\epsilon \sqcup \epsilon_1 \sqcup \epsilon_2)
  }
  \vgap
  \infrule[t-op$^\mathbb{1}$]{
    \Gamma \vdash^s t_1 : \Keywd{nat} \mid \epsilon_1
    \\
    \Gamma \vdash^s t_2 : \Keywd{nat} \mid \epsilon_2
  }{
    \Gamma \vdash^s t_1 \oplus^\mathbb{1} t_2 : \Keywd{nat} \mid (\epsilon_1 \sqcup \epsilon_2)
  }
  \vgap
  \infrule[t-ifz$^\mathbb{1}$]{
    \Gamma \vdash^s t_1 : \Keywd{nat} \mid \epsilon_1
    \\
    \Gamma \vdash^s t_2 : \tau \mid \epsilon_2
    \quad
    \Gamma \vdash^s t_3 : \tau \mid \epsilon_3
  }{
    \Gamma \vdash^s \Keywd{ifz}^\mathbb{1} ~ t_1 ~ t_2 ~ t_3 : \tau \mid (\epsilon_1 \sqcup \epsilon_2 \sqcup \epsilon_3)
  }
  \vgap
  \infrule[t-fix$^\mathbb{1}$]{
    \epsilon_1 = \epsilon_1 \sqcup \epsilon_2
    \quad
    \quad \tau = \tau_1 \to^{\epsilon_1} \tau_2
    \\
    \Gamma \vdash^s t : \tau \to^{\epsilon_2} \tau \mid \epsilon_3
  }{
    \Gamma \vdash^s \Keywd{fix}^\mathbb{1} ~ t : \tau \mid \epsilon_3
  }
  \vgap
  \infrule[t-let]{
    \Gamma \vdash^s t_1 : \tau_1 \mid \epsilon_1
    \\
    \Gamma , x^s : \tau_1 \vdash^s t_2 : \tau_2 \mid \epsilon_2
    \quad
    \WF^s ~ \tau_1
  }{
    \Gamma \vdash^s \Keywd{let} ~ x = t_1 ~ \Keywd{in} ~ t_2 : \tau_2 \mid (\epsilon_1 \sqcup \epsilon_2)
  }
\end{minipage}
\begin{minipage}[t]{.49\linewidth}
  \infrule[t-lam]{
    \Gamma , x^s : \tau_1 \vdash^s t : \tau_2 \mid \epsilon
    \quad
    \WF^s ~ \tau_1
  }{
    \Gamma \vdash^s \lambda x. t : \tau_1 \to^{\epsilon} \tau_2 \mid \bot
  }
  \vgap
  \infrule[t-nat$^\uparrow$]{
    \Gamma \vdash^\mathbb{1} t : \Keywd{nat} \mid \epsilon
  }{
    \Gamma \vdash^\mathbb{1} \Keywd{lift} ~ t : \tyfrag{\Keywd{nat}} \mid \top
  }
  \vgap
  \infrule[t-app$^\mathbb{2}$]{
    \Gamma \vdash^\mathbb{1} t_1 : \tyfrag{(\tau_1 \to^{\bot} \tau_2)} \mid \epsilon_1
    \\
    \Gamma \vdash^\mathbb{1} t_2 : \tyfrag{\tau_1} \mid \epsilon_2
  }{
    \Gamma \vdash^\mathbb{1} \Keywd{app}^\mathbb{2} ~ t_1 ~ t_2 : \tyfrag{\tau_2} \mid \top
  }
  \vgap
  \infrule[t-op$^\mathbb{2}$]{
    \Gamma \vdash^\mathbb{1} t_1 : \tyfrag{\Keywd{nat}} \mid \epsilon_1
    \\
    \Gamma \vdash^\mathbb{1} t_2 : \tyfrag{\Keywd{nat}} \mid \epsilon_2
  }{
    \Gamma \vdash^\mathbb{1} t_1 \oplus^\mathbb{2} t_2 : \tyfrag{\Keywd{nat}} \mid \top
  }
  \vgap
  \infrule[t-ifz$^\mathbb{2}$]{
    \Gamma \vdash^\mathbb{1} t_1 : \tyfrag{\Keywd{nat}} \mid \epsilon_1
    \\
    \Gamma \vdash t_2 : \tyrep{\tau} \mid \epsilon_2
    \quad
    \Gamma \vdash t_3 : \tyrep{\tau} \mid \epsilon_3
  }{
    \Gamma \vdash^\mathbb{1} \Keywd{ifz}^\mathbb{2} ~ t_1 ~ t_2 ~ t_3 : \tyfrag{\tau} \mid \top
  }
  \vgap
  \infrule[t-fix$^\mathbb{2}$]{
    \tau = \tau_1 \to^{\bot} \tau_2
    \\
    \Gamma \vdash^\mathbb{1} t : \tyfrag{(\tau \to^\bot \tau)} \mid \epsilon
  }{
    \Gamma \vdash^\mathbb{1} \Keywd{fix}^\mathbb{2} ~ t : \tyfrag{\tau} \mid \top
  }
  \vgap
  \infrule[t-run]{
    \varnothing \vdash t : \tyrep{\tau} \mid \epsilon
  }{
    \Gamma \vdash^\mathbb{1} \tmrun{t} : \tau \mid \bot
  }
\end{minipage}\\[1ex]
\noindent\rule{13cm}{0.4pt}\\[1ex]
\begin{minipage}[t]{.49\linewidth}
  \infrule[t-pure]{
    \Gamma \vdash^\mathbb{1} t : \tau \mid \bot
  }{
    \Gamma \vdash t : \tau \mid \bot
  }
\end{minipage}
\begin{minipage}[t]{.49\linewidth}
  \infrule[t-rep]{
    \Gamma \vdash^\mathbb{1} t : \tyfrag{\tau} \mid \epsilon
  }{
    \Gamma \vdash t : \tyrep{\tau} \mid \epsilon
  }
\end{minipage}
\caption{Typing rules for \lang's surface syntax.}
\label{fig:typing-pure}
\vspace{-1.5em}
\end{figure}

\begin{figure}[t]
\judgement{Typing Judgments for Administrative Terms}{ \BOX{\Gamma \vdash^s g : \tau \mid \epsilon}}\\[1ex]
\begin{minipage}[t]{.33\linewidth}
  \infrule[t-code-x]{
    x : \tau^{\mathbb{2}} \in \Gamma
    \quad
    \WF^{\mathbb{2}} ~ \tau
  }{
    \Gamma \vdash^\mathbb{1} \Keywd{code} ~ x : \tyfrag{\tau} \mid \bot
  }
\end{minipage}
\begin{minipage}[t]{.33\linewidth}
  \infrule[t-code]{
    \Gamma \vdash^\mathbb{2} t : \tau \mid \bot
  }{
    \Gamma \vdash^\mathbb{1} \Keywd{code} ~ t : \tyrep{\tau} \mid \bot
  }
\end{minipage}
\begin{minipage}[t]{.32\linewidth}
  \infrule[t-refl]{
    \Gamma \vdash^\mathbb{2} t : \tau \mid \bot
  }{
    \Gamma \vdash^\mathbb{1} \Keywd{reflect} ~ t : \tyfrag{\tau} \mid \top
  }
  \vgap
\end{minipage}
\begin{minipage}[t]{.49\linewidth}
  \infrule[t-let$_\mathsf{c}$]{
    \Gamma \vdash^\mathbb{2} t_1 : \tau_1 \mid \bot
    \\
    \Gamma , x^{\mathbb{2}} : \tau_1 \vdash t_2 : \tyrep{\tau_2} \mid \epsilon
    \quad
    \WF^{\mathbb{2}} ~ \tau_1
  }{
    \Gamma \vdash^\mathbb{1} \Keywd{let_c} ~ x = t_1 ~ \Keywd{in} ~ t_2 : \tyrep{\tau_2} \mid \bot
  }
\end{minipage}
\begin{minipage}[t]{.49\linewidth}
  \infrule[t-lam$_\mathsf{c}$]{
    \Gamma, x : \tau_1^{\mathbb{2}} \vdash t : \tyrep{\tau_2} \mid \epsilon
    \quad
    \WF^{\mathbb{2}} ~ \tau_1
  }{
    \Gamma \vdash^\mathbb{1} \lambda_c x. t : \tyfrag{(\tau_1 \to^{\bot} \tau_2)} \mid \top
  }
\end{minipage}
\caption{Typing rules for \lang's administrative terms.}
\label{fig:typing-admin}
\vspace{-2em}
\end{figure}

\lang features a lightweight type-and-effect system that ensures type-safe,
semantics-preserving manipulation of code fragments.
\Cref{fig:static-syntax-pure} presents the syntax of types, effects, and typing contexts.
Types $\tau$ include base types (unit and numbers), function types
$\tau_1 \to^{\epsilon} \tau_2$ with latent effect $\epsilon$, and code types
$\tyrep{\tau}$ and $\tyfrag{\tau}$.
Typing contexts $\Gamma$ map variables $x$ to types $\tau$ and stage annotations $s$.

\subsubsection*{\textbf{Types and Effects for Code Fragments}}
We distinguish two code types: \tyrep{\tau} for complete code values and
\tyfrag{\tau} for fragments not yet reified.
In \lang (and \langref),
general code values may represent effectful computations (\eg, divergence and states),
therefore to preserve their semantics (\Cref{sec:lr}), they must not be
arbitrarily reordered, copied, or discarded.
In contrast, a variable referring to a reified code introduced by
let-insertion (\ie \tmcode{x}) is pure, and can be used unrestrictedly.
Exploiting this observation, our key insight is to assign the unrestricted type
\tyfrag{\tau} both to the pure code reference \tmcode{x} and to the terms that
let-insertion will later be replaced with \tmcode{x}.
Complete code values of type \tyrep{\tau} may contain arbitrary second-stage
effects, and cannot be bound to variables.
Moreover, code fragments of type \tyfrag{\tau} can only appear in specific
contexts that guarantee they will eventually be reified into code values before
being run.

Code generation constructs (\eg, $\mathsf{reflect}$ and $\mathsf{app}^\mathbb{2}$)
behave as a delimited control effect: they trigger reification, and the
surrounding $R$ frame delimits where that effect is handled.
Outside of the $R$ context, the reification effect within the $R$ context is not
observable.
Therefore, in addition to the type distinction, we track if let-insertion could happen
in a term using an effect system.
Effect labels $\epsilon$ indicate whether a term may perform code generation and
let-insertion effects ($\top$) or not ($\bot$).  Effects naturally form a
lattice where $\bot \sqsubseteq \top$, and $\epsilon_1 \sqcup \epsilon_2 =
\epsilon_2$ iff $\epsilon_1 \sqsubseteq \epsilon_2$.

\subsubsection*{\textbf{Binding-Time Well-Formedness}}
\Cref{fig:type-wf-pure} defines stage-indexed type well-formedness,
which the typing rules use to constrain variable stages.
Base types $\mathsf{nat}$ and $\mathsf{unit}$ are well formed at both stages.
A function type $\tau_1 \to^{\epsilon} \tau_2$ is well formed at stage $\mathbb{1}$
if both argument and result types are;
at stage $\mathbb{2}$, both must be well formed at $\mathbb{2}$ and
no further code generation is allowed in the function body ($\epsilon = \bot$).
A code fragment type $\tyfrag{\tau}$ is well formed only at stage $\mathbb{1}$,
with $\tau$ well formed at $\mathbb{2}$.
In contrast, $\tyrep{\tau}$ is \emph{ill-formed at both stages} (thus no corresponding rules),
since allowing such code values (other than \tmcode{x}) to be bound would enable code
duplication or discarding.

\subsubsection*{\textbf{Typing Surface Terms}}

There are two mutually recursive typing judgments: the top-level judgment $\Gamma \vdash
t : \tau \mid \epsilon$ and binding-time sensitive judgment $\Gamma \vdash^s t :
\tau \mid \epsilon$.
The additional binding-time $s$ indicates that term $t$ is checked at stage
$s$. Since \lang is a two-stage language, clearly some terms can only be
checked at the first stage, including $\tmlift{t}$, $\tmrun{t}$, all
$\mathbb{2}$-terms, and all administrative terms.
\looseness=-1

\Cref{fig:typing-pure} presents the typing rules for \lang's surface terms.
Most cases for $\mathbb{1}$-terms are standard.
Most rules for $\mathbb{2}$-terms assign $\mathsf{frag}$-types,
and require the sub-terms to have $\mathsf{frag}$-types too;
they also have the $\top$ effect, indicating the term will be reflected and reified by
let-insertion later.

In the following, we discuss several interesting rules.
Rule $\textsc{t-lam}^\uparrow$ specifies that lifting a term $t$ of first-stage
function type whose argument and return types are both code fragments yields a
second-stage function.
Both $t$'s effect $\epsilon_2$ and the function's latent effect $\epsilon_1$ are
materialized, thus the resulting second-stage function has no
latent let-insertion effect and the effect of the lifting is $\top$.
Rule \textsc{t-run} requires $t$ to be closed, since \lang does not support cross-stage persistence.

The first-stage fixpoint operator (\textsc{t-fix$^\mathbb{1}$}) requires that
the actual recursive function's latent effect $\epsilon_1$ is the upper bound
effect, since we substitute the fixed function itself into the body.
The second-stage conditional expression (\textsc{t-ifz$^\mathbb{2}$}) allows the
branches to be of general code type \tyrep{\tau}, which can be either
terms of $\mathsf{frag}$-type via rule \textsc{t-rep}, or other code
constructs from administrative terms (\Cref{fig:typing-admin}) explained below.

\subsubsection*{\textbf{Typing Administrative Terms}}
\Cref{fig:typing-admin} shows the additional typing rules for administrative terms.
Code variables $\tmcode{x}$ introduced by let-insertion are pure, so they can be
typed as \tyfrag{\tau} (\textsc{t-code-x}) or generally \tyrep{\tau} (\textsc{t-code}).
Rule \textsc{t-code-x} requires the variable $x$ to be bound at stage $\mathbb{2}$ with $\tau$
well-formed at $\mathbb{2}$.

Reflection (\textsc{t-refl}) and $\lambda_\mathsf{c}$-terms (\textsc{t-lamc})
are intermediate reduction constructs eventually reified via let-insertion,
therefore are typed as \tyfrag{\tau} with effect $\top$.
The function body of a $\lambda_\mathsf{c}$-term may have arbitrary effects,
which are materialized by evaluation (under an R context), therefore the resulting
second-stage function has no latent effect.
For $\mathsf{let_c}$-bindings (\textsc{t-let$_\mathsf{c}$}), the right-hand side
$t_1$ is a reified code and is typed under $\vdash^\mathbb{2}$. The body $t_2$ may
perform further first-stage evaluation and requires the general code type \tyrep{\tau_2}.

\subsubsection*{\textbf{Running Example: Ruling Out Stuck Terms}}

Recall from \Cref{sec:keyideas} that programs such as
\lstinline[language=OCaml]|let x = print$^\mathbb{2}$ "Hello" in 42| are ill-typed.
To see this, consider the typing derivation:
the surface typing derives $\mathsf{nat} \mid \top$ at stage
$\mathbb{1}$, but neither \textsc{t-pure} nor \textsc{t-rep} applies for
complete programs:

\newlength{\jw}\newlength{\jwB}
\settowidth{\jw}{$\displaystyle
  \Gamma \vdash^\mathbb{1} \tmlet{x}{\mathsf{print}^{\mathbb{2}}\,\text{``Hello''}}{42} : \mathsf{nat} \mid \top$}
\settowidth{\jwB}{$\displaystyle
  \Gamma \vdash \tmlet{x}{\mathsf{print}^{\mathbb{2}}\,\text{``Hello''}}{42} : \tau \mid \epsilon ~~ \textcolor{red}{\lightning}$}
\ifdim\jwB>\jw \setlength{\jw}{\jwB}\fi
\[
\hspace*{-50pt}
\dfrac{
  \begin{array}{@{}c@{}}
    \Gamma \vdash^\mathbb{1} \mathsf{print}^{\mathbb{2}}\,\text{``Hello''} : \tyfrag{\mathsf{unit}} \mid \top
    \\
    \Gamma, x^\mathbb{1} : \tyfrag{\mathsf{unit}} \vdash^\mathbb{1} 42 : \mathsf{nat} \mid \bot
    \qquad
    \WF^{\mathbb{1}}~\tyfrag{\mathsf{unit}}
  \end{array}
}{
  \dotfrac{
    \Gamma \vdash^\mathbb{1} \tmlet{x}{\mathsf{print}^{\mathbb{2}}\,\text{``Hello''}}{42} : \mathsf{nat} \mid \top
  }{
    \makebox[\jw]{$\displaystyle\Gamma \vdash \tmlet{x}{\mathsf{print}^{\mathbb{2}}\,\text{``Hello''}}{42} : \tau \mid \epsilon$ ~~
      \textcolor{red}{\ding{56}}}
  }%
  \rlap{$\!$\ \footnotesize(Neither \textsc{t-pure} nor \textsc{t-rep} applies \textcolor{red}{\ding{56}})
  }
}
\rlap{$\!$\ \footnotesize(\textsc{t-let})}
\]
Rule \textsc{t-pure} requires effect $\bot$, while \textsc{t-rep} requires an
answer type of the form $\tyfrag{\tau}$. Thus, \lang rejects the program,
preventing evaluation from producing a \code{let$_c$} with a non-code body,
on which the reduction semantics would get stuck.

\subsection{Metatheory: Syntactic Type Soundness} \label{sec:syntactic-soundness}

The \lang-calculus enjoys the syntactic type soundness
properties of progress and preservation \cite{DBLP:journals/iandc/WrightF94}.

\subsubsection*{\textbf{Progress}}
Our type system is crucial for progress: many terms can get stuck in the
reduction semantics, \eg $\tmletc{x}{t}{0}$ when $t$ is not a code value.
As in typical staging calculi, $\lang$ allows evaluation under future-stage
binders, \eg, $\lambda_\mathsf{c} x. \square$ and $\mathsf{let_c} ~ x = t ~
\Keywd{in} ~ \square$ are proper evaluation contexts in \lang.
Therefore, we need to prove a strengthened progress theorem
under general typing contexts including $\mathbb{2}$-bindings.

\begin{lemma}[Strengthened Progress]
\label{lemma:strengthened_progress}
If\: $\Gamma \vdash t : \tau \mid \epsilon$ and
$\forall (x^s : \tau) \in \Gamma.\ s = \mathbb{2}$,
then either $t$ is a value $v$, or there exists $t'$ such that $t \to t'$.
\end{lemma}

\begin{theorem}[Progress]
\label{theorem:progress}
If\: $\varnothing \vdash t : \tau \mid \epsilon$, then $t$ is a value or $\exists t'$ s.t.\ $t \to t'$.
\end{theorem}

\subsubsection*{\textbf{Preservation}}
To show type preservation, we must additionally prove that lifting
$\lambda$-terms (\textsc{st-lam}$^\uparrow$) and let-insertion steps
(\textsc{st-reflect}) preserve typing.
Preservation also states that effects are preserved: evaluation cannot introduce
new effects.
For the lifting of $\lambda$-terms, we prove a substitution lemma that
handles substituting a free variable with an open value
that may contain the same free variable.
The standard substitution lemma still applies for the beta reduction case.
\begin{lemma}[Open Substitution]
\label{lemma:substitution}
If\: $\Gamma , x^{\mathbb{1}} : \tau_1 \vdash^\mathbb{1} t : \tau_2 \mid \epsilon$ and
$\Gamma, x^{s} : \tau_3 \vdash^\mathbb{1} v : \tau_1 \mid \bot$, then
$\Gamma, x^{s} : \tau_3 \vdash^\mathbb{1} t[v/x] : \tau_2 \mid \epsilon$.
\end{lemma}

\begin{theorem}[Preservation]
\label{theorem:preservation}
If\: $\varnothing \vdash t : \tau \mid \epsilon$ and $t \to t'$, then $\exists \epsilon' \sqsubseteq \epsilon$ s.t.\ $\varnothing \vdash t' : \tau \mid \epsilon'$.
\end{theorem}

\section{Erasure and Syntactic Erasure Soundness} \label{sec:erasure}

Recall that our goal is to ensure that the generated code from staging
can be used as a drop-in replacement for the original unstaged program.
The programmer begins with a single-stage program and introduces staging
annotations where specialization is beneficial for performance.
Whether code generation occurs at compile time or run time, the resulting code
should remain extensional equivalent to the original program.
In this section, we formally introduce the notion of erasure
that connects the staged program and its unstaged counterpart.

\subsubsection*{\textbf{Term, Type, and Environment Erasure}}

\begin{figure}
\judgement{Term, Type, and Environment Erasure}{ \BOX{ |t| = t' } \BOX{ |\tau| = \tau' } \BOX{ |\Gamma| = \Gamma' }}\\[0.1em]
\begin{minipage}[t]{0.36\linewidth}
\small
$$
\begin{array}{rclr}
|\texttt{()}| & = & \texttt{()}                                                                     & \\
|n| & = & n                                                                                         & \\
|x| & = & x                                                                                         & \\
|\Keywd{lift} ~ t| & = & |t|                                                                        & \\
|\Keywd{run} ~ t| & = & |t|                                                                         & \\
|\Keywd{app}^s ~ t_1 ~ t_2| & = & \Keywd{app}^\mathbb{1} ~ |t_1| ~ |t_2|                            & \\
|\Keywd{let} ~ x = t_1 ~ \Keywd{in} ~ t_2| & = & \Keywd{let} ~ x = |t_1| ~ \Keywd{in} ~ |t_2|       & \\
|\Keywd{let_c} ~ x = t_1 ~ \Keywd{in} ~ t_2| & = & \Keywd{let} ~ x = |t_1| ~ \Keywd{in} ~ |t_2|     &
\end{array}
$$
\end{minipage}
\hfill
\begin{minipage}[t]{0.29\linewidth}
\small
$$
\begin{array}{rclr}
|\Keywd{fix}^s ~ t| & = & \Keywd{fix}^\mathbb{1} ~ |t|                                              & \\
|\Keywd{ifz}^s ~ t_1 ~ t_2 ~ t_3| & = & \Keywd{ifz}^\mathbb{1} ~ |t_1| ~ |t_2| ~ |t_3|              & \\
|t_1 \oplus^s t_2| & = & |t_1| \oplus^\mathbb{1} |t_2|                                              & \\
|\Keywd{code} ~ t| & = & |t|                                                                        & \\
|\Keywd{reflect} ~ t| & = & |t|                                                                     & \\
|\lambda x. t| & = & \lambda x. |t|                                                                 & \\
|\lambda_\mathsf{c} x. t| & = & \lambda x. |t|                                                      & \\
\end{array}
$$
\end{minipage}
\hfill
\begin{minipage}[t]{0.30\linewidth}
\small
$$
\begin{array}{rclr}
|\Keywd{unit}| & = & \Keywd{unit}                                                                   & \\
|\Keywd{nat}| & = & \Keywd{nat}                                                                     & \\
|\tau_1 \to^{\epsilon} \tau_2| & = & |\tau_1| \to^{\bot} |\tau_2|                                   & \\
|\tyrep{\tau}| & = & |\tau|                                                                         & \\
|\tyfrag{\tau}| & = & |\tau|                                                                        & \\
\hline
|\varnothing| & = & \varnothing                                                                     & \\
|\Gamma, x^s : \tau| & = & |\Gamma|, x^\mathbb{2} : |\tau|                                          &
\end{array}
$$
\end{minipage}
\vspace{-0.5em}
\caption{Term, type, and environment erasure for \lang.}
\label{fig:erasure-all-pure}
\vspace{-1em}
\end{figure}

In \Cref{fig:erasure-all-pure}, we show the definition of term, type, and environment erasure.
Term erasure $|t|$ converts all staging annotations
$s$ in term to $\mathbb{1}$.  Additionally, it strips administrative terms such
as $\Keywd{code}$, $\Keywd{reflect}$, and $\lambda_\mathsf{c}$, which are
generated during staging.  For example, $\Keywd{code} ~ t$ is erased to $|t|$.
Inserted $\Keywd{let_c}$-bindings are transformed to standard $\Keywd{let}$-bindings.

Type erasure $|\tau|$ recursively removes code-related constructors $\mathsf{rep}$ and $\mathsf{frag}$ from types.

Since erased terms no longer induce reification effects,
all latent effects in function types are replaced with $\bot$.
Environment erasure $|\Gamma|$ applies type erasure to all items in the
environment, and sets the binding-time to $\mathbb{2}$.
Clearly, terms and types after erasure are a proper subset of those in \lang.
To avoid confusion, we refer to the language after erasure as a separate
language \langsnd.

\subsubsection*{\textbf{Typing Erased Terms}}
Typings of \langsnd-terms are admissible under \lang's typing rules.
We can precisely characterize the typing of erased terms using
$\Gamma \vdash^\mathbb{2} t : \tau \mid \bot$, where $\Gamma$, $t$, and $\tau$ are all
obtained by erasure.
Intuitively, we are typing terms that coincide with the generated code.
The \textsc{t-code} rule enforces exactly this property, requiring that the code
body to be well-typed as $\Gamma \vdash^\mathbb{2} t : \tau \mid \bot$.
This typing derivation also restricts staging operators from appearing in the
program and enforces the absence of reification effects.

Now we show that erasure preserves syntactic typing:

\begin{theorem}[Syntactic Erasure Soundness]
\label{theorem:syntactic_erasure_soundness}
$\Gamma \vdash t : \tau \mid \epsilon$ implies $|\Gamma| \vdash^\mathbb{2} |t| : |\tau| \mid \bot$.
\end{theorem}
\noindent
Reductions of \langsnd-terms are also admissible under \lang's reduction relation
$t \to t'$, whereas
rules such as \textsc{st-reflect} cannot occur, since erased terms
contain no $\mathbb{2}$-constructs or administrative staging constructs.
When defining the logical relation, we do not distinguish between reductions in \langsnd and \lang;
instead, the binding-time sensitive typing ($\vdash^\mathbb{2}$) regulates which steps
can occur.

\section{Metatheory: Semantics Preservation} \label{sec:lr}

With erasure defined, we now develop the metatheory to
show that staged evaluation in \lang is semantics preserving.
The central task is to show that generated code is contextually equivalent to
its erased counterpart.
To this end, we use step-indexed binary logical relations
\cite{amalthesis,DBLP:conf/popl/AhmedDR09} capturing contextual equivalence
\cite{DBLP:conf/ppdp/BentonKBH07,DBLP:conf/icfp/ThamsborgB11}.
We first introduce the necessary definitions of contextual equivalence, then
build the logical relation, and finally prove that the multi-stage evaluation
of \lang preserves semantics at every reduction step.

\subsection{Contextual Approximation and Equivalence\texorpdfstring{ of \langsnd}{}}

\begin{figure}[!htb]
  \begin{flushleft}
  \judgement{Observational Context for Contextual Equivalence}\\[1ex]
  \end{flushleft}
$$
\begin{array}{rcll}
\textsf{Frame} & F & ::= &
  \lambda x. \square
  \mid \mathsf{let} ~ x = \square ~ \Keywd{in} ~ t
  \mid \mathsf{let} ~ x = t ~ \Keywd{in} ~ \square
  \mid \mathsf{app}^\mathbb{1} ~ \square ~ t
  \mid \mathsf{app}^\mathbb{1} ~ t ~ \square
  \mid \mathsf{fix}^\mathbb{1} ~ \square \\
  & & \hspace{0.5em} \mid &
  \mathsf{ifz}^\mathbb{1} ~ \square ~ t_2 ~ t_3
  \mid \mathsf{ifz}^\mathbb{1} ~ t_1 ~ \square ~ t_3
  \mid \mathsf{ifz}^\mathbb{1} ~ t_1 ~ t_2 ~ \square
  \mid \square ~ \oplus^\mathbb{1} ~ t_2
  \mid t_1 ~ \oplus^\mathbb{1} ~ \square \\
\textsf{Contexts} & C & ::= & \square \mid C \circ F \\
\end{array}
$$

\judgement{Context Typing Rules}
{
    \BOX{F : (\Gamma \vdash \tau) \Rightarrow (\Gamma' \vdash \tau')}
    \BOX{C : (\Gamma \vdash \tau) \Rightarrow (\Gamma' \vdash \tau')}
}

\vgap

\begin{minipage}[t]{.64\linewidth}
  \infrule[f-lam]{
    \WF^\mathbb{2} ~ \tau_1
  }
  {
    \lambda x. \square :
    (\Gamma, x^\mathbb{2} : \tau_1 \vdash \tau_2) \Rightarrow
    (\Gamma \vdash \tau_1 \to^{\bot} \tau_2)
  }
\end{minipage}

\vgap

\begin{minipage}[t]{.40\linewidth}
  \infrule[f-let-1]{
    \Gamma, x^\mathbb{2} : \tau_1 \vdash^\mathbb{2} t : \tau_2 \mid \bot
  }
  {
    \mathsf{let} ~ x = \square ~ \Keywd{in} ~ t :
    (\Gamma \vdash \tau_1) \Rightarrow
    (\Gamma \vdash \tau_2)
  }
\end{minipage}
\begin{minipage}[t]{.40\linewidth}
  \infrule[f-let-2]{
    \Gamma \vdash^\mathbb{2} t : \tau_1 \mid \bot
  }
  {
    \mathsf{let} ~ x = t ~ \Keywd{in} ~ \square :
    (\Gamma, x^\mathbb{2} : \tau_1 \vdash \tau_2) \Rightarrow
    (\Gamma \vdash \tau_2)
  }
\end{minipage}

\vgap

\begin{minipage}[t]{.40\linewidth}
  \infrule[f-app-1]{
    \Gamma \vdash^\mathbb{2} t : \tau_1 \mid \bot
  }
  {
    \mathsf{app}^\mathbb{1} ~ \square ~ t :
    (\Gamma \vdash \tau_1 \to^{\bot} \tau_2) \Rightarrow
    (\Gamma \vdash \tau_2)
  }
\end{minipage}
\begin{minipage}[t]{.40\linewidth}
  \infrule[f-app-2]{
    \Gamma \vdash^\mathbb{2} t : \tau_1 \to^{\bot} \tau_2 \mid \bot
  }
  {
    \mathsf{app}^\mathbb{1} ~ t ~ \square :
    (\Gamma \vdash \tau_1) \Rightarrow
    (\Gamma \vdash \tau_2)
  }
\end{minipage}

\vgap

\begin{minipage}[t]{.40\linewidth}
  \infrule[f-fix]{
    \tau = \tau_1 \to^{\bot} \tau_2
  }
  {
    \mathsf{fix}^\mathbb{1} ~ \square :
    (\Gamma \vdash \tau \to^{\bot} \tau) \Rightarrow
    (\Gamma \vdash \tau)
  }
\end{minipage}
\begin{minipage}[t]{.40\linewidth}
  \infrule[f-ifz-1]{
    \Gamma \vdash^\mathbb{2} t_2 : \tau \mid \bot
    \quad
    \Gamma \vdash^\mathbb{2} t_3 : \tau \mid \bot
  }
  {
    \mathsf{ifz}^\mathbb{1} ~ \square ~ t_2 ~ t_3 :
    (\Gamma \vdash \Keywd{nat}) \Rightarrow
    (\Gamma \vdash \tau)
  }
\end{minipage}

\vgap

\begin{minipage}[t]{.40\linewidth}
  \infrule[f-ifz-2]{
    \Gamma \vdash^\mathbb{2} t_1 : \Keywd{nat} \mid \bot
    \quad
    \Gamma \vdash^\mathbb{2} t_3 : \tau \mid \bot
  }
  {
    \mathsf{ifz}^\mathbb{1} ~ t_1 ~ \square ~ t_3 :
    (\Gamma \vdash \tau) \Rightarrow
    (\Gamma \vdash \tau)
  }
\end{minipage}
\begin{minipage}[t]{.40\linewidth}
  \infrule[f-ifz-3]{
    \Gamma \vdash^\mathbb{2} t_1 : \Keywd{nat} \mid \bot
    \quad
    \Gamma \vdash^\mathbb{2} t_2 : \tau \mid \bot
  }
  {
    \mathsf{ifz}^\mathbb{1} ~ t_1 ~ t_2 ~ \square :
    (\Gamma \vdash \tau) \Rightarrow
    (\Gamma \vdash \tau)
  }
\end{minipage}

\vgap

\begin{minipage}[t]{.40\linewidth}
  \infrule[f-op-1]{
    \Gamma \vdash^\mathbb{2} t_2 : \Keywd{nat} \mid \bot
  }
  {
    \square ~ \oplus^\mathbb{1} ~ t_2 :
    (\Gamma \vdash \Keywd{nat}) \Rightarrow
    (\Gamma \vdash \Keywd{nat})
  }
\end{minipage}
\begin{minipage}[t]{.40\linewidth}
  \infrule[f-op-2]{
    \Gamma \vdash^\mathbb{2} t_1 : \Keywd{nat} \mid \bot
  }
  {
    t_1 ~ \oplus^\mathbb{1} ~ \square :
    (\Gamma \vdash \Keywd{nat}) \Rightarrow
    (\Gamma \vdash \Keywd{nat})
  }
\end{minipage}

\vgap

\begin{minipage}[t]{.40\linewidth}
  \infrule[C-hole]{
  }
  {
    \square :
    (\Gamma \vdash \tau) \Rightarrow
    (\Gamma \vdash \tau)
  }
\end{minipage}
\begin{minipage}[t]{.40\linewidth}
  \infrule[C-cons]{
    C : (\Gamma' \vdash \tau') \Rightarrow (\Gamma'' \vdash \tau'')
    \\
    F : (\Gamma \vdash \tau) \Rightarrow (\Gamma' \vdash \tau')
  }
  {
    C \circ F :
    (\Gamma \vdash \tau) \Rightarrow
    (\Gamma'' \vdash \tau'')
  }
\end{minipage}

\caption{Context typing rules for \langsnd.}
\label{fig:ctx-equiv-pure}
\end{figure}

Two programs are contextually equivalent iff no well-formed observational
context distinguishes them.
Since both stage-erased programs and generated programs belong to \langsnd,
an observational context $C$ is a partial \langsnd program with hole
$\square$.
We write $C : (\Gamma \vdash \tau) \Rightarrow (\Gamma' \vdash \tau')$ if
$\Gamma \vdash^\mathbb{2} t : \tau \mid \bot$ entails
$\Gamma' \vdash^\mathbb{2} C[t] : \tau' \mid \bot$.
\Cref{fig:ctx-equiv-pure} defines observational contexts and their typing rules.

Because \langsnd programs may diverge, we define contextual equivalence in
terms of termination: two terms are indistinguishable when they either both
terminate or both diverge.
In particular, if two terminating terms produce different values,
there must exist an
observational context that can distinguish them by making one terminate and the
other diverge.

Following the standard approach in the literature \cite{DBLP:conf/popl/AhmedDR09}, we factor
contextual equivalence into two directions of contextual approximation, \ie,
termination of one program implies termination of the other.
Contextual equivalence is their conjunction.
Because the reduction relation is deterministic
(\Cref{theorem:deterministic}), contextual equivalence is transitive.

\begin{definition}[Contextual Approximation]
If $\Gamma \vdash^\mathbb{2} t_1 : \tau \mid \bot$, and $\Gamma \vdash^\mathbb{2} t_2 : \tau \mid \bot$, and
$ \forall C : (\Gamma \vdash \tau) \Rightarrow (\varnothing \vdash \tau').\, C[t_1]\Downarrow\ \Longrightarrow C[t_2]\Downarrow$,
then $t_1$ contextually approximates $t_2$,
written as $\Gamma \vDash t_1 \preceq^{\text{ctx}} t_2 : \tau$.
We write $t \Downarrow$ to denote that $t$ terminates, \ie, $t \to^* v$ for some value $v$.
\end{definition}

\begin{definition}[Contextual Equiv.]
$
\Gamma \vDash t_1 \simeq^{\text{ctx}} t_2 : \tau \stackrel{\text{def}}{=}
\Gamma \vDash t_1 \preceq^{\text{ctx}} t_2 : \tau ~ \wedge ~
\Gamma \vDash t_2 \preceq^{\text{ctx}} t_1 : \tau.
$
\end{definition}

\begin{theorem}[Transitivity]
\label{theorem:transitivity}
If\:
$\Gamma \vDash t_1 \simeq^{\text{ctx}} t_2 : \tau$ and
$\Gamma \vDash t_2 \simeq^{\text{ctx}} t_3 : \tau$, then
$\Gamma \vDash t_1 \simeq^{\text{ctx}} t_3 : \tau$.
\looseness=-1
\end{theorem}

\subsection{Binary Logical Relations}

\begin{figure}
\resizebox{\textwidth}{!}{
$
\begin{array}{rclr}

\mathcal{V}\llbracket   \Keywd{unit}               \rrbracket & = & \{ (k, \texttt{()}, \texttt{()}) \} & \\
\mathcal{V}\llbracket   \Keywd{nat}                \rrbracket & = & \{ (k, n, n) \} & \\
\mathcal{V}\llbracket   \tau_1 \to^{\bot} \tau_2   \rrbracket & = &
    \{ (k, \lambda x. t_1, \lambda x. t_2) \mid
    \forall j \le k.\ \forall v_1, v_2.\ (j, v_1, v_2) \in \mathcal{V}\llbracket \tau_1 \rrbracket \Rightarrow \\
    & &\hspace{1em} (j, \tmappSt{(\lambda x. t_1)}{v_1}, \tmappSt{(\lambda x. t_2)}{v_2}) \in \mathcal{E}\llbracket \tau_2 \rrbracket
    \} &

\\[1ex]

\mathcal{E}\llbracket   \tau    \rrbracket & = &
    \{ (k, t_1, t_2) \mid
    \forall j < k.\ \forall v_1.\ t_1 \to^j v_1 \Rightarrow \\
    & &\hspace{1em} \exists v_2.\
    t_2 \to^* v_2 ~ \wedge ~
    (k - j, v_1, v_2) \in \mathcal{V}\llbracket \tau \rrbracket
    \} &

\\[1ex]

\mathcal{G}\llbracket \varnothing \rrbracket & = & \{ (k, \emptyset, \emptyset) \} & \\
\mathcal{G}\llbracket \Gamma, x^\mathbb{2} : \tau \rrbracket & = &
    \{ (k, \gamma_1[x \mapsto v_1], \gamma_2[x \mapsto v_2]) \mid
        (k, \gamma_1, \gamma_2) \in \mathcal{G}\llbracket \Gamma \rrbracket ~ \wedge ~
        (k, v_1, v_2) \in \mathcal{V}\llbracket \tau \rrbracket
    \}
    &

\\[1ex]

\Gamma \vDash t_1 \preceq^{\text{log}} t_2 : \tau & \stackrel{\text{def}}{=} &
    \Gamma \vdash^\mathbb{2} t_1 : \tau \mid \bot ~ \wedge ~
    \Gamma \vdash^\mathbb{2} t_2 : \tau \mid \bot ~ \wedge ~ \\
    & & \forall k \ge 0.\ \forall \gamma_1, \gamma_2.~
    (k, \gamma_1, \gamma_2) \in \mathcal{G}\llbracket \Gamma \rrbracket \Rightarrow
    (k, \gamma_1(t_1), \gamma_2(t_2)) \in \mathcal{E}\llbracket \tau \rrbracket
    & \\

\Gamma \vDash t_1 \simeq^{\text{log}} t_2 : \tau & \stackrel{\text{def}}{=} &
    \Gamma \vDash t_1 \preceq^{\text{log}} t_2 : \tau ~ \wedge ~
    \Gamma \vDash t_2 \preceq^{\text{log}} t_1 : \tau
\end{array}
$
}
\vspace{-1em}
\caption{The binary interpretation of types and terms, and logical equivalence for \langsnd.}
\label{fig:binary_lr_pure}
\vspace{-1em}
\end{figure}

\Cref{fig:binary_lr_pure} defines the binary logical relation.
Because both erased terms and generated code admit divergence, the
relation is step-indexed to be well-founded, following prior work
\cite{amalthesis,DBLP:conf/popl/AhmedDR09}.

The value relation is given by the semantic interpretation of types, written
$\mathcal{V}\llbracket \tau \rrbracket$.
For a type $\tau$, it consists of triples $(k, v_1, v_2)$, where $k$ is the
step index (\ie, a natural number) and $v_1$, $v_2$ are closed values.
Unit and numbers are related iff they are syntactically equal.
Two $\lambda$-terms are related if applying them to related arguments at a
future step index yields related results.
\looseness=-1

The term relation $\mathcal{E}\llbracket \tau \rrbracket$ is likewise defined
semantically by type.
Intuitively, $(k, t_1, t_2) \in \mathcal{E}\llbracket \tau \rrbracket$ means
that whenever $t_1$ reduces to a value $v_1$ in $j < k$ steps, $t_2$ reduces to
some value $v_2$ such that
$(k - j, v_1, v_2) \in \mathcal{V}\llbracket \tau \rrbracket$.
The environment relation $\mathcal{G}\llbracket \Gamma \rrbracket$ is defined
inductively on the structure of type environments.
In the inductive case, two substitutions are related if they map $x$ to values
related at type $\tau$, extending substitutions $\gamma_1$ and $\gamma_2$ that
are already related.

The logical approximation $\preceq^{\text{log}}$ relates two open terms
under $\Gamma$ if substituting their free variables with
related substitutions $\gamma_1$ and $\gamma_2$ yields related closed terms.
We write $\gamma(t)$ for the substitution of each free variable in $t$ by its
corresponding value in $\gamma$.
Logical equivalence between $t_1$ and $t_2$ is then defined as approximation in
both directions.

\subsubsection*{\textbf{Soundness of Logical Relations}}
The compatibility lemmas are standard and omitted for brevity; they are the
semantic counterparts of a subset of the typing rules (\Cref{fig:typing-pure})
and can be found in the mechanized proof.
These lemmas yield the fundamental theorem: a syntactically well-typed term is
logically related to itself.

\begin{theorem}[Fundamental Property]
\label{theorem:fundamental}
If\:
$\Gamma \vdash^\mathbb{2} t : \tau \mid \bot$, then
$\Gamma \vDash t \simeq^{\text{log}} t : \tau$.
\end{theorem}

\noindent
The following soundness theorem is our license to use the logical relation to prove contextual equivalence.
Moreover, it serves as an important tool in proving the semantics preservation of staging,
as contextual equivalence is often difficult to use directly in proofs.

\begin{theorem}[Soundness of Logical Relations]
\label{theorem:soundness_of_logical_relations}
If\:
$\Gamma \vDash t_1 \simeq^{\text{log}} t_2 : \tau$, then
$\Gamma \vDash t_1 \simeq^{\text{ctx}} t_2 : \tau$.
\end{theorem}

\subsection{Semantics Preservation of Staging}

\subsubsection*{\textbf{Overview}}
We now state the central meta-theorem of \lang.
If a two-stage program $t_1$ evaluates to $\tmcode{t_2}$, then the generated
code $t_2$ is contextually equivalent to the unstaged program $|t_1|$ obtained
by erasing staging annotations.
This equivalence is captured by the logical relations above.
Thus, well-typed staging annotations neither introduce nor eliminate behaviors
of the original unstaged program.
This guarantee is strictly stronger than syntactic erasure soundness
(\Cref{theorem:syntactic_erasure_soundness}), and we therefore call it
\emph{semantics preservation} of staging.

To prove this result, we first establish a stronger invariant: semantics are
preserved by \emph{every} reduction step, including those that produce administrative
terms.
\Cref{theorem:single_step_semantic_preservation} states that a redex and its
contractum are contextually equivalent after erasure, with the key lemma
showing that let-insertion preserves evaluation order.
Here we rely on the soundness of the logical relation
(\Cref{theorem:soundness_of_logical_relations}) to derive contextual
equivalence.
From this single-step result, multi-step preservation follows.
Combined with transitivity of contextual equivalence,
we can prove that evaluation to a code value preserves semantics
(\Cref{theorem:semantic_preservation}).
\Cref{fig:proof} illustrates the proof structure.
\looseness=-1

\subsubsection*{\textbf{Semantics Preservation of Single-Step Reduction}}
We now turn to the main lemmas.
They show that substitution (for ordinary $\mathsf{let}$ and
$\beta$-reduction) and let-insertion both preserve evaluation order after
erasure; the remaining reduction cases are routine.

We discuss the substitution case for
$\Keywd{let}~x=v~\Keywd{in}~t$ (\textsc{st-let}); the application case is
analogous.
For let-insertion, the term $E[\Keywd{reflect}~t]$ steps under a $P$-context to
$\Keywd{let_c}~x=t~\Keywd{in}~E[\Keywd{code}~x]$.
This preserves evaluation order after erasure because $|E|$ remains an
evaluation context,
since $E$-contexts are restricted to \emph{pure} frames.
We define erasure on contexts $|E|$ directly from term erasure.
By contrast, reification contexts do not in general erase to evaluation
contexts: for example, $\lambda_\mathsf{c} x.\square$ erases to
$\lambda x.\square$, but the erased language does not evaluate under
$\lambda$-binders.

\begin{lemma}[Semantics Preservation of Substitution]
\label{lemma:substitution_semantics_soundness}
Substitution preserves evaluation order: if
$\Gamma \vdash \Keywd{let} ~ x = v ~ \Keywd{in} ~ t : \tau \mid \epsilon$, then
$|\Gamma| \vDash |\Keywd{let} ~ x = v ~ \Keywd{in} ~ t| \simeq^{\text{ctx}} |t[v/x]| : |\tau|$.
\end{lemma}

\begin{lemma}[Semantics Preservation of Let-insertion]
\label{lemma:let_insertion_semantics_soundness}
Let-insertion preserves evaluation order: if
$\ \Gamma \vdash E[\Keywd{reflect} ~ t] : \tau \mid \epsilon$, then
$\ |\Gamma| \vDash |E[\Keywd{reflect} ~ t]| \simeq^{\text{ctx}} |\Keywd{let_c} ~ x = t ~ \Keywd{in} ~ E[\Keywd{code} ~ x]| : |\tau|$.
\end{lemma}

\noindent
With these lemmas, we prove that a single reduction step preserves semantics:

\begin{theorem}[Semantics Preservation of Single-Step Reduction]
\label{theorem:single_step_semantic_preservation}
If $\ \Gamma \vdash t_1 : \tau \mid \epsilon$, and
$t_1 \to t_2$, then
$|\Gamma| \vDash |t_1| \simeq^{\text{ctx}} |t_2| : |\tau|$.
\end{theorem}

\subsubsection*{\textbf{Semantics Preservation of Staging}}
Finally, we present the central theorem:
if a two-stage program $t_1$ is typed with $\tyrep{\tau}$,
and $t_1$ generates a program $\tmcode{t_2}$,
then $|t_1|$ is contextually equivalent to $t_2$.
It suffices to show a stronger theorem relating an arbitrary redex and its
contractum, which gives us a strengthened inductive hypothesis.
The proof relies on the transitivity of contextual equivalence to extend
single-step semantics preservation to multi-steps.

\begin{theorem}[Strengthened Semantics Preservation]
\label{theorem:strengthened_semantic_preservation}
If\: $\Gamma \vdash t_1 : \tau \mid \epsilon$ and $t_1 \to^* t_2$, then
$|\Gamma| \vDash |t_1| \simeq^{\text{ctx}} |t_2| : |\tau|$.
\end{theorem}

\begin{theorem}[Semantics Preservation]
\label{theorem:semantic_preservation}
If\: $\varnothing \vdash t_1 : \tyrep{\tau} \mid \epsilon$ and $t_1 \to^* \tmcode{t_2}$, then
$\varnothing \vDash |t_1| \simeq^{\text{ctx}} t_2 : \tau$.
\end{theorem}

\subsubsection*{\textbf{Remark}}
Our results provide a precise semantic account of let-insertion in staging and
answer the question posed by \citet{DBLP:conf/esop/InoueT12,DBLP:journals/jfp/InoueT16}:
``\emph{When does erasing staging annotations preserve semantics?}''
For the CBV $\lang$-calculus, our answer is simple and definitive:
erasing annotations in \lang \emph{always} preserves semantics.

\section{Extension to Mutable References} \label{sec:mutref}

We next extend \lang with mutable references, yielding \langref, to study
erasure soundness for stateful staged programs.
We first characterize the interactions between staging and references that
break semantic preservation.
We then adopt a simple, practical design that
confines store effects to the second stage while keeping the first stage
store-pure.
The type system enforces this separation while allowing generated code to
allocate and update references, as in the staged power function from
\Cref{sec:motivation}. This section presents the syntax, semantics, and
metatheory of \langref; \Cref{sec:extension} discusses possible designs
supporting first-stage store effects.

\subsection{When Do Store Effects Break Erasure Soundness?} \label{sec:when-break}

Suppose we extend \lang with first-order mutable references. Certain
interactions between staging and references can then violate erasure soundness.
Consider a function \code{inc} that captures a first-stage reference \code{r}, initialized
to \code{0}, and increments it; the program ultimately reads \code{r}. The
staged program appears on the left and its stage-erased counterpart on the
right.

\noindent
\begin{minipage}{.52\textwidth}
\begin{lstlisting}[language=OCaml]
  (* staged version *)
  let r = ref 0 in (* r 1st-stage ref *)
  let inc = lift (fun y -> r := !r+1; y) in
  lift (!r)
\end{lstlisting}
\end{minipage}%
\hfill\vline\hfill
\begin{minipage}{.47\textwidth}
\begin{lstlisting}[language=OCaml]
 (* stage-erased version *)
 let r = ref 0 in
 let inc = (fun y -> r := !r+1; y) in
 !r
\end{lstlisting}
\end{minipage}
In the staged program, \code{r} is allocated in the first stage.
When the function is lifted, however, the assignment
\code{r := !r+1} is performed during first-stage evaluation, even though the
function is never applied.
As a result, the final expression \code{lift (!r)} residualizes the value
\code{1}, and the generated code is essentially
\code{let inc = (fun y -> y) in 1}.
By contrast,
in the stage-erased program, the body of \code{inc} remains under
the $\lambda$ and is never called.
Therefore, the increment does not occur, and the final value of \code{r}
remains \code{0}.

The discrepancy shows that
lifting a function forces static effects in its body to occur during first-stage
evaluation, whereas after erasure those same effects remain latent
and are performed only when the function is applied.
Because function bodies are not evaluation contexts after erasure, the two
programs no longer exhibit the same effects, and erasure soundness breaks.

The same problem also arises with second-stage conditionals.
Consider the following example, in which both branches access the same
reference.
The then-branch mutates the reference before reading it, whereas the else-branch
reads it directly.

\noindent
\begin{minipage}{.52\textwidth}
\begin{lstlisting}[language=OCaml]
  (* staged version *)
  let r = ref 0 in (* r 1st-stage ref *)
  if$^\mathbb{2}$ (lift b)
    then { r := !r + 1; lift (!r) }
    else lift (!r)
\end{lstlisting}
\end{minipage}%
\hfill\vline\hfill
\begin{minipage}{.47\textwidth}
\begin{lstlisting}[language=OCaml]
 (* stage-erased version *)
 let r = ref 0 in
 if b
   then { r := !r + 1; !r }
   else !r
\end{lstlisting}
\end{minipage}

In the stage-erased program, only one branch and its effects are evaluated.
In the staged program, by contrast, residualizing the conditional
requires reifying both branches, along with the first-stage effects
performed while reifying them.
As a result, the generated code may depend on the order in
which the branches are reified.
In our calculi, the then-branch is reified first.
Thus, when the else-branch is reified, it observes that
\code{r} has already been incremented to \code{1}.
Consequently, the generated code is \code{if b then 1 else 1},
whereas the stage-erased program evaluates to either \code{1} or \code{0}.

Developing a principled solution that rules out such counterexamples is
non-trivial and beyond the scope of this paper; see \Cref{sec:extension} for
discussion.
The examples above, however, suggest that residualizing the allocation of
\code{r} to the second stage is benign for semantics preservation.
We adopt this discipline in \langref: although it restricts certain use cases, it
still permits meaningful manipulation of effectful code fragments, and serves as
a stepping stone toward future expressive designs.

\subsection{The Syntax, Semantics, and Erasure of \texorpdfstring{\langref}{}} \label{sec:langref-syntax}

This section describes the extensions that \langref adds to \lang.
\Cref{fig:langref-extension} presents the new syntax, reduction semantics,
and type system, with changed rules relative to \lang highlighted in red.
Similar to \lang, operations on mutable references are
attached with a stage annotation (\eg, $\mathsf{alloc}^s$).
Locations $\ell$ are administrative terms and values, which do not appear in source terms.
We restrict to first-order stores for examining interactions between staging and references, therefore
store $\sigma$ maps locations to numbers.

\subsubsection*{\textbf{Dynamic Semantics}}
The reduction semantics with stores is mostly standard.
We extend the reduction relation of \lang with a runtime store,
$\langle \sigma, ~t \rangle \to \langle \sigma', ~t' \rangle$.
Because the type system confines store effects to the second stage,
first-stage evaluation always uses the empty store:
$\langle \varnothing, ~t \rangle \to^* \langle \varnothing, ~v \rangle$.
Nevertheless, when $v = \tmcode{t'}$, the generated code is then evaluated
in the second stage with actual store effects,
$\langle \varnothing, ~t' \rangle \to^* \langle \sigma, ~v' \rangle$,
yielding the final value.
In this way, first-stage evaluation may still manipulate effectful code
fragments which later allocate or mutate the store.

Pure frames $B$ are extended for effectful operations.
As in \lang, the new rules for effectful $\mathbb{2}$-terms leave the
store unchanged and instead reflect second-stage effectful terms
into corresponding first-stage terms.
The new rules for effectful $\mathbb{1}$-terms have the standard form
$\sigma, ~t \headto \sigma', ~t'$
(\Cref{fig:langref-extension}), including \textsc{st-alloc}$^\mathbb{1}$,
\textsc{st-get}$^\mathbb{1}$, and \textsc{st-put}$^\mathbb{1}$.
These rules may occur during second-stage evaluation under any
evaluation context $M$ (\textsc{st-mut}).
Finally, \textsc{st-pure} and \textsc{st-reflect} are extended to
thread the store through configurations without changing it.

\subsubsection*{\textbf{Static Semantics}}

\begin{figure}
\judgement{Syntax}{}
\[
\begin{array}{rcl@{\qquad}rcl}
\textsf{Terms} ~ t & ::= & \ldots
\mid \Keywd{alloc}^s ~ t
\mid \Keywd{get}^s ~ t
\mid \Keywd{put}^s ~ t_1 ~ t_2
&
\textsf{Locations} ~ \ell & \in & \mathbb{N} \\[2pt]
\textsf{Administrative Terms} ~ g & ::= & \ldots \mid \ell
&
\textsf{Values} ~ v & ::= & \ldots \mid \ell \\[2pt]
\textsf{Store} ~ \sigma & ::= & \varnothing \mid \sigma[\ell \mapsto n]
&
\textsf{Types} ~ \tau & ::= & \ldots \mid \Keywd{ref} ~ \tau \\[2pt]
\textsf{Pure Frames} ~ B & ::= & \multicolumn{4}{l}{\ldots
  \mid \mathsf{alloc}^s ~ \square
  \mid \mathsf{get}^s ~ \square
  \mid \mathsf{put}^s ~ \square ~ t
  \mid \mathsf{put}^s ~ v ~ \square}
\end{array}
\]

\vgap

\judgement{Dynamic Semantics}{ \BOX{t \headto t'} \BOX{\sigma, ~t \headto \sigma', ~t'} \BOX{\textcolor{red!70!black}{\langle} \textcolor{red!70!black}{\sigma}\textcolor{red!70!black}{,} ~t \textcolor{red!70!black}{\rangle} \to \textcolor{red!70!black}{\langle} \textcolor{red!70!black}{\sigma'}\textcolor{red!70!black}{,} ~t' \textcolor{red!70!black}{\rangle}}}
\[
\begin{array}{lrcll}
\steprule{st-alloc$^\mathbb{2}$} & \tmallocDy{(\tmcode{t})} & \headto & \tmreflect{(\tmallocSt{t})} & \\
\steprule{st-get$^\mathbb{2}$} & \tmgetDy{(\tmcode{t})} & \headto & \tmreflect{(\tmgetSt{t})} & \\
\steprule{st-put$^\mathbb{2}$} & \tmputDy{(\tmcode{t_1})}{(\tmcode{t_2})} & \headto & \tmreflect{(\tmputSt{t_1}{t_2})} & \\
\steprule{st-alloc$^\mathbb{1}$} & \sigma, ~\tmallocSt{n} & \headto & \sigma[\ell \mapsto n], ~\ell \hspace{3em} \text{where}~ \ell \notin \mathit{DOM}(\sigma) & \\
\steprule{st-get$^\mathbb{1}$} & \sigma, ~\tmgetSt{\ell} & \headto & \sigma, ~n  \hspace{6em} \text{where}~ \sigma(\ell) = n & \\
\steprule{st-put$^\mathbb{1}$} & \sigma, ~\tmputSt{\ell}{n} & \headto & \sigma[\ell \mapsto n], ~\texttt{()} &
\end{array}
\]

\begin{minipage}[t]{.40\linewidth}
  \infrule[st-pure]{
    t \headto t'
  }{
    \textcolor{red!70!black}{\langle} \textcolor{red!70!black}{\sigma}\textcolor{red!70!black}{,} ~M[t] \textcolor{red!70!black}{\rangle} \to \textcolor{red!70!black}{\langle} \textcolor{red!70!black}{\sigma}\textcolor{red!70!black}{,} ~M[t'] \textcolor{red!70!black}{\rangle}
  }
\end{minipage}
\begin{minipage}[t]{.49\linewidth}
  \infrule[st-mut]{
    \sigma, ~t \headto \sigma', ~t'
  }{
    \langle \sigma, ~M[t] \rangle \to \langle \sigma', ~M[t'] \rangle
  }
\end{minipage}\\[1ex]
\begin{minipage}[t]{.7\linewidth}
  \infrule[st-reflect]{
    x~\text{is~fresh}
  }{\
    \textcolor{red!70!black}{\langle} \textcolor{red!70!black}{\sigma}\textcolor{red!70!black}{,} ~P[E[\tmreflect{t}]] \textcolor{red!70!black}{\rangle} \to \textcolor{red!70!black}{\langle} \textcolor{red!70!black}{\sigma}\textcolor{red!70!black}{,} ~P[\tmletc{x}{t}{E[\tmcode{x}]}] \textcolor{red!70!black}{\rangle}
  }
\end{minipage}

\vgap

\begin{flushleft}
\judgement{Static Semantics}{}
\vspace{0.5ex}
\end{flushleft}
\begin{minipage}[t]{.45\linewidth}
  \infrule[wf-ref]{
    \WF^{s} ~ \tau
  }{
    \WF^{s} ~ (\Keywd{ref} ~ \tau)
  }
  \vgap
  \infrule[t-alloc$^\mathbb{1}$]{
    \Gamma \vdash^\mathbb{2} t : \Keywd{nat} \mid \epsilon
  }{
    \Gamma \vdash^\mathbb{2} \Keywd{alloc}^\mathbb{1} ~ t : \Keywd{ref} ~ \Keywd{nat} \mid \epsilon
  }
  \vgap
  \infrule[t-get$^\mathbb{1}$]{
    \Gamma \vdash^\mathbb{2} t : \Keywd{ref} ~ \mathsf{nat} \mid \epsilon
  }{
    \Gamma \vdash^\mathbb{2} \Keywd{get}^\mathbb{1} ~ t : \mathsf{nat} \mid \epsilon
  }
  \vgap
  \infrule[t-put$^\mathbb{1}$]{
    \Gamma \vdash^\mathbb{2} t_1 : \Keywd{ref} ~ \Keywd{nat} \mid \epsilon_1
    \quad
    \Gamma \vdash^\mathbb{2} t_2 : \Keywd{nat} \mid \epsilon_2
  }{
    \Gamma \vdash^\mathbb{2} \Keywd{put}^\mathbb{1} ~ t_1 ~ t_2 : \Keywd{unit} \mid (\epsilon_1 \sqcup \epsilon_2)
  }
\end{minipage}
\begin{minipage}[t]{.53\linewidth}
  \infrule[t-run]{
    \textcolor{red!70!black}{\mathsf{storeFree}~t} \quad
    \varnothing \vdash t : \tyrep{\tau} \mid \epsilon
  }{
    \Gamma \vdash^\mathbb{1} \tmrun{t} : \tau \mid \bot
  }
  \vgap
  \infrule[t-alloc$^\mathbb{2}$]{
    \Gamma \vdash^\mathbb{1} t : \tyfrag{\Keywd{nat}} \mid \epsilon
  }{
    \Gamma \vdash^\mathbb{1} \Keywd{alloc}^\mathbb{2} ~ t : \tyfrag{(\Keywd{ref} ~ \Keywd{nat})} \mid \top
  }
  \vgap
  \infrule[t-get$^\mathbb{2}$]{
    \Gamma \vdash^\mathbb{1} t : \tyfrag{(\Keywd{ref} ~ \mathsf{nat})} \mid \epsilon
  }{
    \Gamma \vdash^\mathbb{1} \Keywd{get}^\mathbb{2} ~ t : \tyfrag{\mathsf{nat}} \mid \top
  }
  \vgap
  \infrule[t-put$^\mathbb{2}$]{
    \Gamma \vdash^\mathbb{1} t_1 : \tyfrag{(\Keywd{ref} ~ \Keywd{nat})} \mid \epsilon_1
    \quad
    \Gamma \vdash^\mathbb{1} t_2 : \tyfrag{\Keywd{nat}} \mid \epsilon_2
  }{
    \Gamma \vdash^\mathbb{1} \Keywd{put}^\mathbb{2} ~ t_1 ~ t_2 : \tyfrag{\Keywd{unit}} \mid \top
  }
\end{minipage}\\[1ex]
\vspace{-1em}
\caption{The syntax and semantics of \langref. Only extended 
formalizations on top of \lang are included.}
\label{fig:langref-extension}
\vspace{-1em}
\end{figure}

\langref extends the types with mutable references, written $\Keywd{ref}~\tau$,
whose well-formedness is standard: $\tau$ must be well formed at the same stage.
The typing rules for second-stage effectful operations
($\Keywd{alloc}^\mathbb{2}$, $\Keywd{get}^\mathbb{2}$, and $\Keywd{put}^\mathbb{2}$)
follow the existing second-stage rules of \lang.
However, to restrict store effects only to the second stage,
the first-stage effectful operations
($\mathsf{alloc}^\mathbb{1}$, $\mathsf{get}^\mathbb{1}$, and $\mathsf{put}^\mathbb{1}$)
are allowed only under second-stage typing judgments ($\vdash^\mathbb{2}$), where they appear solely as generated code.
The \textsf{run} operation must be restricted as well, as running code
that contains store effects could introduce locations back into the first stage.
Thus, rule \textsc{t-run} requires $t$ to be store-effect free,
\ie, it contains no
$\mathsf{alloc}^s$, $\mathsf{get}^s$, or $\mathsf{put}^s$ operations;
the full definition of $\mathsf{storeFree}$ can be found in
\Cref{sec:storefree}.

Although \langref includes references and locations,
we do not need store or location typing for the sake of
proving syntactic type soundness of first-stage evaluation and
semantics preservation.
Since first-stage evaluation has no store effects,
the rules in \Cref{fig:langref-extension} are sufficient to prove
syntactic type soundness.
For the logical relation, we instead use a stronger world model for
annotation-erased terms and generated code, which likewise avoids
syntactic store typing.

The erasure operation on new terms and types are straightforward
and structurally defined, similar to \Cref{fig:erasure-all-pure};
they are omitted and can be found in the mechanization.

\subsection{Type Soundness and Semantics Preservation for \texorpdfstring{\langref}{}} \label{sec:langref-sem}

\begin{figure}
\resizebox{\textwidth}{!}{
$
\begin{array}{rclr}
W' \sqsupseteq W           & \stackrel{\text{def}}{=} & \forall (\ell_1, \ell_2) \in W. (\ell_1, \ell_2) \in W' & \\
(j, W') \sqsupseteq (k, W) & \stackrel{\text{def}}{=} & j \le k \wedge W' \sqsupseteq W & \\
(\sigma_1, \sigma_2) : W   & \stackrel{\text{def}}{=} & \forall (\ell_1, \ell_2) \in W. \sigma_1(\ell_1) = \sigma_2(\ell_2) &
\\[1ex]

\mathcal{V}\llbracket   \Keywd{unit}               \rrbracket & = & \{ (k, W, \texttt{()}, \texttt{()}) \} & \\
\mathcal{V}\llbracket   \Keywd{nat}                \rrbracket & = & \{ (k, W, n, n) \} & \\
\mathcal{V}\llbracket   \Keywd{\Keywd{ref} ~ nat}  \rrbracket & = & \{ (k, W, \ell_1, \ell_2) \mid (\ell_1, \ell_2) \in W \} & \\
\mathcal{V}\llbracket   \tau_1 \to^{\bot} \tau_2   \rrbracket & = &
    \{ (k, W, \lambda x. t_1, \lambda x. t_2) \mid
    \forall (j, W') \sqsupseteq (k, W).\ \forall v_1, v_2.\ (j, W', v_1, v_2) \in \mathcal{V}\llbracket \tau_1 \rrbracket \Rightarrow \\
    & &\hspace{1em} (j, W', \tmappSt{(\lambda x. t_1)}{v_1}, \tmappSt{(\lambda x. t_2)}{v_2}) \in \mathcal{E}\llbracket \tau_2 \rrbracket
    \} &

\\[1ex]

\mathcal{E}\llbracket   \tau    \rrbracket & = &
    \{ (k, W, t_1, t_2) \mid
    \forall j < k.\ \forall \sigma_1, \sigma_2, \sigma_1', v_1.\
    (\sigma_1, \sigma_2) : W ~ \wedge ~ \langle \sigma_1, t_1 \rangle \to^j \langle \sigma_1', v_1 \rangle \Rightarrow \\
    & &\hspace{1em} \exists \sigma_2', v_2, W'.\
    \langle \sigma_2, t_2 \rangle \to^* \langle \sigma_2', v_2 \rangle ~ \wedge ~
    (k - j, W') \sqsupseteq (k, W) ~ \wedge ~ \\
    & &\hspace{1em} (\sigma_1', \sigma_2') : W' ~ \wedge ~
    (k - j, W', v_1, v_2) \in \mathcal{V}\llbracket \tau \rrbracket
    \} &

\\[1ex]

\mathcal{G}\llbracket \varnothing \rrbracket & = & \{ (k, W, \emptyset, \emptyset) \} & \\
\mathcal{G}\llbracket \Gamma, x^\mathbb{2} : \tau \rrbracket & = &
    \{ (k, W, \gamma_1[x \mapsto v_1], \gamma_2[x \mapsto v_2]) \mid
        (k, W, \gamma_1, \gamma_2) \in \mathcal{G}\llbracket \Gamma \rrbracket ~ \wedge ~
        (k, W, v_1, v_2) \in \mathcal{V}\llbracket \tau \rrbracket
    \}
    &

\\[1ex]

\Gamma \vDash t_1 \preceq^{\text{log}} t_2 : \tau & \stackrel{\text{def}}{=} &
    \Gamma \vdash^\mathbb{2} t_1 : \tau \mid \bot ~ \wedge ~
    \Gamma \vdash^\mathbb{2} t_2 : \tau \mid \bot ~ \wedge ~ \\
    & & \forall k \ge 0.\ \forall \gamma_1, \gamma_2, W.~
    (k, W, \gamma_1, \gamma_2) \in \mathcal{G}\llbracket \Gamma \rrbracket \Rightarrow
    (k, W, \gamma_1(t_1), \gamma_2(t_2)) \in \mathcal{E}\llbracket \tau \rrbracket
    & \\

\Gamma \vDash t_1 \simeq^{\text{log}} t_2 : \tau & \stackrel{\text{def}}{=} &
    \Gamma \vDash t_1 \preceq^{\text{log}} t_2 : \tau ~ \wedge ~
    \Gamma \vDash t_2 \preceq^{\text{log}} t_1 : \tau
\end{array}
$
}
\caption{The world model, binary logical relations, and logical equivalence definitions for \langsndref.}
\label{fig:binary_lr}
\vspace{-1em}
\end{figure}

The \langref-calculus enjoys all meta-theoretical properties as \lang,
including syntactic type soundness, syntactic erasure soundness, and semantics preservation.
Syntactic type soundness is standard, as the type system ensures that
first-stage evaluation is store-free.
\begin{theorem}[Progress]
\label{theorem:ref-progress}
If $~\vdash t : \tau \mid \epsilon$, then $t$ is a value or $\exists t'$ s.t.\ $\langle \varnothing, t \rangle \to \langle \varnothing, t' \rangle$.
\end{theorem}
\vspace{-0.5em}
\begin{theorem}[Preservation]
\label{theorem:ref-preservation}
If $~\vdash t : \tau \mid \epsilon$ and $\langle \varnothing, t \rangle \to \langle \varnothing, t' \rangle$,
then $\exists \epsilon' \sqsubseteq \epsilon$ s.t.\ $\varnothing \vdash t' : \tau \mid \epsilon'$.
\end{theorem}
\vspace{-0.5em}
\begin{theorem}[Syntactic Erasure Soundness]
\label{theorem:ref-syn-erasure-soundness}
If\: $\Gamma \vdash t : \tau \mid \epsilon$, then $|\Gamma| \vdash^\mathbb{2} |t| : |\tau| \mid \bot$.
\end{theorem}

Below we outline the extension of the logical relation
to relate stores via a world model, and then sketch our approach to semantics
preservation.

\subsubsection*{\textbf{The World Model}}

Unlike \lang that has no store and uses step-indexed logical relations alone,
\langref must relate stores across two runs.
We introduce worlds $W \subseteq \mathbb{N} \times \mathbb{N}$ to
relate locations in two stores (\Cref{fig:binary_lr}).
Because \langref restricts stores to natural numbers, there are no cyclic references,
thus the world need not be recursively indexed by types and steps.
A world $W$ is a partial bijection on locations (each location paired
with at most one counterpart).
We write $(\ell_1, \ell_2) \in W$ when locations are related, and $(\sigma_1, \sigma_2) : W$ when both stores agree on all related locations.
Worlds are ordered by extension: $W' \sqsupseteq W$ means that $W'$ adds new related locations while preserving relations in $W$.

\subsubsection*{\textbf{Semantics Preservation}}
The logical relation is extended to relate stores via the world model (\Cref{fig:binary_lr}).
Value, term, and environment interpretations are indexed by both the step $k$
and the world $W$.
For example, $(k, W, t_1, t_2) \in \mathcal{E}\llbracket \tau \rrbracket$
relates two terms under
the initial stores $(\sigma_1, \sigma_2) : W$, and
if in $k$ reduction steps we obtain values $v_1, v_2$,
then the values are related in the extended world $W' \sqsupseteq W$.
Logical approximation $\preceq^{\text{log}}$ is defined as in \lang,
but with quantification over all worlds $W$.
The observational contexts and their typing rules are likewise extended
with frames for the effectful operations in \Cref{fig:full-ctx-equiv},
yielding the contextual equivalence $\simeq^{\text{ctx}}$ used in
\Cref{theorem:ref-semantic_preservation}.
We also prove the fundamental property and soundness of the logical relation.

\begin{figure}[!t]
\begin{flushleft}
\judgement{Observational Context for Contextual Equivalence}\\[1ex]
\end{flushleft}
$$
\begin{array}{rcll}
\textsf{Frame} & F & ::= \ldots
  \mid \mathsf{alloc}^\mathbb{1} ~ \square
  \mid \mathsf{get}^\mathbb{1} ~ \square
  \mid \mathsf{put}^\mathbb{1} ~ \square ~ t_2
  \mid \mathsf{put}^\mathbb{1} ~ t_1 ~ \square
\\
\end{array}
$$
\judgement{Context Typing Rules}
{
    \BOX{F : (\Gamma \vdash \tau) \Rightarrow (\Gamma' \vdash \tau')}
}\\[1ex]

\vgap

\begin{minipage}[t]{.40\linewidth}
  \infrule[f-alloc]{
  }
  {
    \mathsf{alloc}^\mathbb{1} ~ \square :
    (\Gamma \vdash \Keywd{nat}) \Rightarrow
    (\Gamma \vdash \Keywd{ref} ~ \Keywd{nat})
  }
\end{minipage}
\begin{minipage}[t]{.40\linewidth}
  \infrule[f-get]{
  }
  {
    \mathsf{get}^\mathbb{1} ~ \square :
    (\Gamma \vdash \Keywd{ref} ~ \Keywd{nat}) \Rightarrow
    (\Gamma \vdash \Keywd{nat})
  }
\end{minipage}

\vgap

\begin{minipage}[t]{.40\linewidth}
  \infrule[f-put-1]{
    \Gamma \vdash^\mathbb{2} t_2 : \Keywd{nat} \mid \bot
  }
  {
    \mathsf{put}^\mathbb{1} ~ \square ~ t_2 :
    (\Gamma \vdash \Keywd{ref} ~ \Keywd{nat}) \Rightarrow
    (\Gamma \vdash \Keywd{unit})
  }
\end{minipage}
\begin{minipage}[t]{.40\linewidth}
  \infrule[f-put-2]{
    \Gamma \vdash^\mathbb{2} t_1 : \Keywd{ref} ~ \Keywd{nat} \mid \bot
  }
  {
    \mathsf{put}^\mathbb{1} ~ t_1 ~ \square :
    (\Gamma \vdash \Keywd{nat}) \Rightarrow
    (\Gamma \vdash \Keywd{unit})
  }
\end{minipage}

\caption{Additional context typing rules for \langsndref.
Only extended formalizations on top of \langsnd are included.}
\label{fig:full-ctx-equiv}
\end{figure}

Finally, we prove semantics preservation with the new step-indexed
Kripke logical relation:
\vspace{-0.5em}
\begin{theorem}[Semantics Preservation]
\label{theorem:ref-semantic_preservation}
If $~\vdash t_1 : \tyrep{\tau} \mid \epsilon$ and $\langle \varnothing, t_1 \rangle \to^* \langle \varnothing, \tmcode{t_2} \rangle$,
then $\varnothing \vDash |t_1| \simeq^{\text{ctx}} t_2 : \tau$.
\end{theorem}
\vspace{-1em}

\subsubsection*{\textbf{Remarks}}
With the \langref-calculus, we show that it is possible to design a
staging calculus with mutable references that enjoys erasure soundness.
We again provide a simple and definitive answer to
\citet{DBLP:conf/esop/InoueT12,DBLP:journals/jfp/InoueT16}'s question:
erasing annotations in \langref \emph{always} preserves semantics.
\looseness=-1

\section{Open Questions and Future Work} \label{sec:extension}

We have presented \lang and \langref, along with their type soundness and semantic
preservation results.
A central open question is whether one can design an expressive multi-stage
language with flexible store effects while still preserving semantics.
As discussed in \Cref{sec:when-break}, the main obstacle is that $\lambda_c$-terms
and 2nd-stage conditionals cease to be evaluation contexts after erasure,
so two-stage evaluation may produce store effects absent from the stage-erased
program.

Still, it is not entirely hopeless.
Whereas \langref enforces \emph{syntactic purity} at the first stage, a more
flexible approach could enforce \emph{semantic purity}
\cite{DBLP:journals/jfp/Sabry98, DBLP:journals/corr/abs-2510-07582}, admitting
store effects that provably do not interfere with staging. For example, lifting
a $\lambda$-term whose local store is fully encapsulated can preserve contextual
equivalence. One promising direction is to adapt Rust-style resource-tracking
type systems \cite{DBLP:journals/toplas/BoruchGruszeckiOLLB23,
DBLP:journals/pacmpl/WeiBJBR24, DBLP:conf/esop/MarshallVO22,
DBLP:journals/pacmpl/BaoWBJHR21, DBLP:journals/pacmpl/XuBO24} to ensure that the
store reachable during first-stage evaluation is disjoint from that accessed by
$\lambda_c$-terms. Such separation could make effects semantically invariant
under reordering, duplication, or omission. Region-based binding-time analyses
in partial evaluation offer a related precedent: \citet{dussart1997partial} use
region information to approximate disjoint portions of the store, suggesting a
basis for identifying non-interfering effects.

For conditionals, one could similarly require the locations reachable from the two
branches to be disjoint, so that the order would not matter.
Another possible approach, at the cost of code duplication, is polyvariant
specialization: branches together with their
shared continuations are specialized under different stores
or more fine-grained stack-like structure of stores \cite{dussart1997partial}.

More than a decade ago, \citet{DBLP:conf/esop/InoueT12} identified erasure
soundness for imperative MSP as an open challenge, observing that ``imperative
MSP is not yet ready.'' Recent advances in resource-tracking type systems bring
flexible, semantics-preserving imperative MSP within reach. Our results provide
a foundation for exploring this rich design space.

\subsubsection*{\textbf{Extensions}}
We discuss several possible extensions to \langref that are left for future work.
\begin{itemize}[leftmargin=1.5em]
  \item \langref models second-stage first-order mutable store for simplicity.
  Extending it to support \emph{second-stage higher-order store} should be
  straightforward following existing logical relations and world models for
  higher-order store \cite{DBLP:conf/popl/AhmedDR09, amalthesis}.
  \item The two-stage setting covers many practical applications and provides a
  foundation for $N$-stage programs ($N>2$). Our binary logical relation could
  serve as the final-stage component, relating the last stage to its generated
  code. Semantic typing \cite{DBLP:journals/jacm/TimanyKDB24} and logical
  relations for code types offer promising paths toward this generalization.
  \item \lang's context decomposition $P[E[\tmreflect{t}]]$ determines a unique,
  semantically sound position for inserting let-bindings. This design represents
  one point in a broader space of code-motion strategies: pure computations
  need not always be let-inserted, and impure computations may be moved as well
  as long as they preserve semantics.
  A finer-grained effect system that tracks dependencies
  \cite{DBLP:journals/pacmpl/BracevacWJAJBR23} could enable more flexible
  insertion strategies.
\end{itemize}

\section{Mechanization in Lean} \label{sec:mechanization}
One of our main contributions is the mechanization of the \lang/\langref
calculi and their metatheory in Lean (v.~4.29.0).
All results are machine-checked.
\Cref{tab:formalization} summarizes the structure and effort of our development,
comprising about 10k lines for $\lang$ and 12k lines for $\langref$.

\begin{table}[t]
\centering
\caption{Summary of the mechanization in Lean.} \label{tab:formalization}
\small
\vspace{-1em}
\scalebox{0.85}{%
\begin{tabular}{l|p{7.2cm}|c|c}
Component             & Description & LoC of $\lang$ & LoC of $\langref$ \\ \hline
Syntax                & Syntax and syntactic operations using locally nameless.            & 2k    & 2.5k  \\
OperationalSemantics  & Static semantics and Determinism.                                  & 2k    & 2.5k  \\
SyntacticTyping       & Dynamic semantics and Syntactic Erasure Soundness.                 & 1k    & 1.1k  \\
SyntacticSoundness    & Preservation and Progress.                                         & 1.5k  & 1.7k  \\
CtxEquiv              & Contextual equivalence.                                            & 0.2k  & 0.3k  \\
LogicalEquiv          & Logical relations and the Fundamental theorem.                     & 1.9k  & 2k    \\
SemanticsPreservation & Single-step and Multi-step Semantic Preservation.                  & 1.4k  & 1.7k  \\
Total LoC             &                                                                    & 10k   & 11.8k \\
\end{tabular}
}
\end{table}

\subsubsection*{\textbf{Locally Nameless with de Bruijn Levels}}
Variables are represented in the locally nameless style
\cite{DBLP:journals/jar/Chargueraud12}: free variables use de Bruijn levels and
bound variables use de Bruijn indices.
Using de Bruijn indices for bound variables simplifies the generation of fresh
variables in let-insertion and avoids $\alpha$-equivalence.
We use de Bruijn levels rather than explicit names for free variables,
which simplifies the introduction of new free variables during typing.

\subsubsection*{\textbf{Level-Indexed Reduction and Contexts}}
Because reification contexts introduce second-stage bindings and can be plugged
into terms in which second-stage variables occur free, our reduction relation is
indexed by the current de Bruijn level.
Evaluation contexts also track the current de Bruijn level, and their levels
inform the reduction relation.
For example, $\lambda_\mathsf{c} x.\, \square$ increases the current level by one
when entering its body.
At the top level, the well-formed levels of reduction and evaluation contexts are 0.

\subsubsection*{\textbf{Deterministic Memory Allocation}}
Without loss of generality, we simplify the design of our store in the
formalization by modeling it as a stack rather than a heap. Our locations start
at 0 and increment continuously with each allocation. This simplification still
ensures that the properties we aim to discuss remain sufficiently strong.

\section{Related Work} \label{sec:related}

\subsubsection*{\textbf{General Staged Programming}}
MetaML-style staging languages \cite{DBLP:conf/pepm/TahaS97, taha1999multistage,
DBLP:conf/flops/Kiselyov14, 10.1145/1809028.1806642} provide syntactic
annotations that can change evaluation strategies.
LMS \cite{DBLP:conf/gpce/RompfO10, DBLP:conf/birthday/Rompf16}, by contrast,
uses host-language operator overloading and does not rely on explicit
syntax for staging. To model LMS-style staging that involves let-insertion,
\citet{DBLP:journals/pacmpl/AminR18} introduced the untyped multi-stage
$\lambda_{\uparrow\downarrow}$-calculus with automatic let-insertion. Building
on this foundation, we specialize the calculus to two stages, develop a sound
type-and-effect system, extend it with mutable references, and prove semantic
preservation under erasure. Our results formalize the conditions under which
let-insertion preserves semantics and clarify its interaction with
effects.
This semantic guarantee also provides a foundation for stage-polymorphic
programming \cite{DBLP:journals/pacmpl/AminR18, DBLP:conf/esop/HengleinM94,
DBLP:conf/gpce/OfenbeckRP17}, enabling programmers to abstract over stages and
write stage-generic code that behaves uniformly across staging choices.

\subsubsection*{\textbf{Let-Insertion in Staging and Specialization}}

Let-insertion has long supported specialization of
effectful programs in partial evaluation \cite{dussart1997partial,
DBLP:conf/popl/Danvy96, DBLP:books/daglib/0072559, DBLP:conf/tacs/LawallT97,
DBLP:journals/mscs/HatcliffD97}.
Modern staging systems realize let-insertion in several ways.
For example, BER MetaOCaml
provides the explicit \code{genlet} primitive~\cite{DBLP:journals/scp/Kiselyov26},
implemented using delimited control \cite{DBLP:journals/tcs/Kiselyov12, DBLP:conf/lfp/DanvyF90}.
Similarly, \citet{DBLP:conf/pepm/KameyamaKS09, DBLP:journals/jfp/KameyamaKS11}
combined delimited control with code generation and let-insertion.
LMS performs let-insertion automatically through its dependency-tracking graph
IR \cite{DBLP:conf/gpce/RompfO10, DBLP:conf/birthday/Rompf16,
DBLP:journals/pacmpl/BracevacWJAJBR23}.
Squid also also performs A-normal form transformation for object code
\cite{DBLP:conf/scala/ParreauxSK17, DBLP:journals/pacmpl/ParreauxVSK18,
DBLP:conf/pldi/FlanaganSDF93}.

Our calculi, like $\lambda_{\uparrow\downarrow}$
\cite{DBLP:journals/pacmpl/AminR18}, model let-insertion through small-step
reduction. \citet{10.1145/3828676} instead present several big-step definitional
interpreters using states, one-CPS, two-CPS, or control operators,
together with an abstract machine for the same flavor of let-insertion.
\lang and \langref require no explicit delimited-control operators; their
type-and-effect systems instead track let-insertion as a control effect.
MetaOCaml similarly distinguishes general code fragments from object-code
variables produced by \code{genletv}.
Our mechanized metatheory formalizes this distinction and establishes its role
in type and erasure soundness.

In principle, MetaML-style staging can simulate behavior similar to our calculi
by inserting \code{genlet} ubiquitously.
However, rather than permitting unconstrained code-generation behavior as in MetaML,
we advocate semantics-preserving staging as a more disciplined and
predictable default.
Programmers can always opt out explicitly when necessary, much like \code{unsafe} in Rust.
This design makes potentially semantics-breaking behavior visible and
intentional, instead of leaving it implicit.
In practice, staging libraries such as LMS take this approach as well,
relying on the host-language Scala's by-name arguments to support code duplication.

\subsubsection*{\textbf{Reasoning Multi-Stage Programs and Partial Evaluation}}

In partial evaluation, \citet{DBLP:books/daglib/0072559} established the
correctness of LambdaMix which supports general recursion, and
\citet{DBLP:conf/ppdp/Filinski99} proved TDPE sound and complete for a pure
call-by-name $\lambda$-calculus.
\citet{DBLP:conf/tacs/LawallT97} proved on-the-fly let-insertion sound, but did
not consider effects performed during specialization, as discussed in \Cref{sec:when-break}.

\citet{yang2000reasoning} was the first to propose the \emph{erasure soundness}
property, and regarded it as a criterion of ``semantic correctness'' for code
generation, complementing the ``syntactic correctness'' guarantee of well-typed
generated code. He proved erasure soundness for a CBN two-level
language without mutable state, and identified duplication of effects as the key
reason why a naive CBV two-level language fails to preserve semantics.
Although let-insertion was proposed as a potential solution, a formal treatment
or proofs were not provided in his work.

\citet{DBLP:journals/jfp/InoueT16, DBLP:conf/esop/InoueT12} shared the same
concern about the correctness of staging and developed equational theories and
applicative bisimulation for the $\lambda^U$-calculus.
They established termination conditions for erasure soundness: under CBV,
it requires both the staged program and its stage-erased counterpart to
terminate at first-order constants.
In their setting, the calculus permits divergence but has no store effects, and
their results posit stronger purity assumptions.
Both \citet{yang2000reasoning} and our work instead assume that first-stage
evaluation terminates, ensuring that object code exists, and then relate that
code to the erased program in the unstaged base language.
In contrast, \citet{DBLP:journals/jfp/InoueT16, DBLP:conf/esop/InoueT12}
studied the equational theory of the staged language itself.

Independently and concurrently with this work, \citet{ct075_agda_staging}
mechanized an LMS-like calculus and proved a similar semantics-preservation
theorem in Agda. Although the top-level theorems are similar, there are several
key differences: First, we use a small-step reduction semantics, whereas
\citet{ct075_agda_staging} follows the big-step stateful semantics of
\citet{DBLP:conf/birthday/Rompf16}. Second, our calculi model both divergence and
store effects, whereas the calculus of \citet{ct075_agda_staging} requires
strong normalization and does not include store effects. Finally, the proof
techniques differ: we use binary Kripke logical relations that do not
directly model code types, whereas \citet{ct075_agda_staging} directly
incorporates the equivalence between a two-stage term and its generated code
into the value-type relation for code types.

\citet{lmcs:929} developed a Hoare logic for reasoning about Mini-ML$^\square_e$.
In a dependently typed setting, \citet{DBLP:journals/pacmpl/Kovacs22} developed
two-level type theory (2LTT), where staging is given by the
evaluation of 2LTT syntax in a semantic domain.  He also proved the
conservativity of 2LTT.

\section{Conclusion}

In this paper, we study the question posed by \citet{DBLP:conf/esop/InoueT12}:
when do staging annotations preserve the semantics of the underlying
single-stage program?
To this end, we design the \lang- and \langref-calculi that
support automatic let-insertion and include a lightweight
type-and-effect system for tracking code-generation effects.
Our main results establish syntactic type soundness via progress and
preservation, and semantics preservation via step-indexed Kripke logical relations.
We believe that the design of these calculi, together with their mechanized
semantics-preservation proofs, fills a long-standing gap in reasoning about
effectful multi-stage programs using modern logical-relation techniques.
We hope this work serves as a solid foundation for future developments in
expressive, reliable, and semantics-preserving multi-stage languages.

\section*{Data-Availability Statement}
Our artifact includes the mechanization of the $\lang$-calculus
and the $\langref$-calculus, as well as examples and their meta-theoretic results in Lean 4.
All theorems in the paper are mechanically proved in the artifact.
The artifact is available at
\url{https://doi.org/10.5281/zenodo.21392703} \cite{artifact_zenodo}
and \url{https://github.com/instar-lang/mechanization/tree/oopsla26}.

\begin{acks}
We thank the reviewers of OOPSLA 2026 and PLDI 2026 for their detailed and constructive comments.
We thank Tiark Rompf, Cameron Wong, and Peter Thiemann for helpful discussions.
\end{acks}

\bibliographystyle{ACM-Reference-Format}
\bibliography{references}

\clearpage
\appendix
\section{The Store-Free Predicate} \label{sec:storefree}

\begin{figure}[!htb]
\judgement{Store-Free Assertion}{ \BOX{\mathsf{storeFree} ~ t}}\\[1ex]
\begin{minipage}[t]{.32\linewidth}
  \infrule[sf-unit]{}
  {
    \mathsf{storeFree} ~ \texttt{()}
  }
\end{minipage}
\begin{minipage}[t]{.32\linewidth}
  \infrule[sf-nat]{}
  {
    \mathsf{storeFree} ~ n
  }
\end{minipage}
\begin{minipage}[t]{.32\linewidth}
  \infrule[sf-var]{}
  {
    \mathsf{storeFree} ~ x
  }
\end{minipage}

\vgap

\begin{minipage}[t]{.32\linewidth}
  \infrule[sf-lam]
  {
    \mathsf{storeFree} ~ t
  }
  {
    \mathsf{storeFree} ~ (\lambda x.t)
  }
\end{minipage}
\begin{minipage}[t]{.32\linewidth}
  \infrule[sf-lift]
  {
    \mathsf{storeFree} ~ t
  }
  {
    \mathsf{storeFree} ~ (\tmlift{t})
  }
\end{minipage}
\begin{minipage}[t]{.32\linewidth}
  \infrule[sf-run]
  {
    \mathsf{storeFree} ~ t
  }
  {
    \mathsf{storeFree} ~ (\tmrun{t})
  }
\end{minipage}

\vgap

\begin{minipage}[t]{.32\linewidth}
  \infrule[sf-lam$_\mathsf{c}$]
  {
    \mathsf{storeFree} ~ t
  }
  {
    \mathsf{storeFree} ~ (\tmlamc{x}{t})
  }
\end{minipage}
\begin{minipage}[t]{.32\linewidth}
  \infrule[sf-code]
  {
    \mathsf{storeFree} ~ t
  }
  {
    \mathsf{storeFree} ~ (\tmcode{t})
  }
\end{minipage}
\begin{minipage}[t]{.32\linewidth}
  \infrule[sf-refl]
  {
    \mathsf{storeFree} ~ t
  }
  {
    \mathsf{storeFree} ~ (\tmreflect{t})
  }
\end{minipage}

\vgap

\begin{minipage}[t]{.48\linewidth}
  \infrule[sf-let$_\mathsf{c}$]
  {
    \mathsf{storeFree} ~ t_1 \quad \mathsf{storeFree} ~ t_2
  }
  {
    \mathsf{storeFree} ~ (\tmletc{x}{t_1}{t_2})
  }
\end{minipage}
\begin{minipage}[t]{.48\linewidth}
  \infrule[sf-fix]
  {
    \mathsf{storeFree} ~ t
  }
  {
    \mathsf{storeFree} ~ (\Keywd{fix}^s ~ t)
  }
\end{minipage}
\vgap

\begin{minipage}[t]{.48\linewidth}
  \infrule[sf-let]
  {
    \mathsf{storeFree} ~ t_1 \quad \mathsf{storeFree} ~ t_2
  }
  {
    \mathsf{storeFree} ~ (\Keywd{let} ~ x = t_1 ~ \Keywd{in} ~ t_2)
  }
\end{minipage}
\begin{minipage}[t]{.48\linewidth}
  \infrule[sf-app]
  {
    \mathsf{storeFree} ~ t_1 \quad \mathsf{storeFree} ~ t_2
  }
  {
    \mathsf{storeFree} ~ (\Keywd{app}^s ~ t_1 ~ t_2)
  }
\end{minipage}

\vgap

\begin{minipage}[t]{.48\linewidth}
  \infrule[sf-op]
  {
    \mathsf{storeFree} ~ t_1 \quad \mathsf{storeFree} ~ t_2
  }
  {
    \mathsf{storeFree} ~ (t_1 \oplus^s t_2)
  }
\end{minipage}
\begin{minipage}[t]{.48\linewidth}
  \infrule[sf-ifz]
  {
    \mathsf{storeFree} ~ t_1 \\
    \mathsf{storeFree} ~ t_2 \quad \mathsf{storeFree} ~ t_3
  }
  {
    \mathsf{storeFree} ~ (\Keywd{ifz}^s ~ t_1 ~ t_2 ~ t_3)
  }
\end{minipage}

\caption{Store-Free assertion in \langref.}
\label{fig:store-free}
\end{figure}

\end{document}